\documentclass[12pt]{article}
\usepackage{amssymb}
\usepackage{epsfig}
\usepackage{bbm}
\usepackage{subfigure}

\usepackage{graphicx}

\def\L{\mathcal L}
\def\e{\varepsilon}

\newcommand{\wt}{\widetilde}
\newcommand{\mathsym}[1]{{}}

\newcommand{\p}{\varphi}

\begin{document}

\def\a{\alpha}
\def\b{\beta}
\def\c{\chi}
\def\d{\delta}
\def\e{\epsilon}
\def\f{\phi}
\def\g{\gamma}
\def\h{\eta}
\def\j{\psi}
\def\k{\kappa}
\def\l{\lambda}
\def\m{\mu}
\def\n{\nu}
\def\o{\omega}
\def\p{\pi}
\def\q{\theta}
\def\r{\rho}
\def\s{\sigma}
\def\t{\tau}
\def\u{\upsilon}
\def\x{\xi}
\def\z{\zeta}
\def\D{\Delta}
\def\F{\Phi}
\def\G{\Gamma}
\def\J{\Psi}
\def\L{\Lambda}
\def\O{\Omega}
\def\P{\Pi}
\def\Q{\Theta}
\def\S{\Sigma}
\def\U{\Upsilon}
\def\X{\Xi}
\def\Rmsma{r}
\def\Rmsmasq{r^2}
\def\ms{m_{\rm sol}}
\def\ma{m_{\rm atm}}
%Varletters
\def\ve{\varepsilon}
\def\vf{\varphi}
\def\vr{\varrho}
\def\vs{\varsigma}
\def\vq{\vartheta}

\def\nn{\nonumber}
\def\dg{\dagger}                                     % hermitian conjugate
\def\ddg{\ddagger}                                   % double dagger
\def\wt#1{\widetilde{#1}}                    % big tilde
\def\mt{\widetilde{m}_1}
\def\mti{\widetilde{m}_i}
\def\mtj{\widetilde{m}_j}
\def\rt{\widetilde{r}_1}
\def\mtt{\widetilde{m}_2}
\def\mttt{\widetilde{m}_3}
\def\rtt{\widetilde{r}_2}
\def\mb{\overline{m}}
\def\VEV#1{\left\langle #1\right\rangle}        % < >
\def\be{\begin{equation}}
\def\ee{\end{equation}}
\def\ds{\displaystyle}
\def\ra{\rightarrow}
\def\dd{\displaystyle}
\def\bea{\begin{eqnarray}}
\def\eea{\end{eqnarray}}
\def\NO{\nonumber}
\def\Bar#1{\overline{#1}}

% ------- Definitions --------
\def\lsim{\ \rlap{\raise 3pt \hbox{$<$}}{\lower 3pt \hbox{$\sim$}}\ }
\def\gsim{\ \rlap{\raise 3pt \hbox{$>$}}{\lower 3pt \hbox{$\sim$}}\ }
\def\cY{{\cal Y}}
\def\hcY{\hat{\cal Y}}

% Journal abbreviations (preprints)

\def\pl#1#2#3{Phys.~Lett.~{\bf B {#1}} ({#2}) #3}
\def\np#1#2#3{Nucl.~Phys.~{\bf B {#1}} ({#2}) #3}
\def\prl#1#2#3{Phys.~Rev.~Lett.~{\bf #1} ({#2}) #3}
\def\pr#1#2#3{Phys.~Rev.~{\bf D {#1}} ({#2}) #3}
\def\zp#1#2#3{Z.~Phys.~{\bf C {#1}} ({#2}) #3}
\def\cqg#1#2#3{Class.~and Quantum Grav.~{\bf {#1}} ({#2}) #3}
\def\cmp#1#2#3{Commun.~Math.~Phys.~{\bf {#1}} ({#2}) #3}
\def\jmp#1#2#3{J.~Math.~Phys.~{\bf {#1}} ({#2}) #3}
\def\ap#1#2#3{Ann.~of Phys.~{\bf {#1}} ({#2}) #3}
\def\prep#1#2#3{Phys.~Rep.~{\bf {#1}C} ({#2}) #3}
\def\ptp#1#2#3{Progr.~Theor.~Phys.~{\bf {#1}} ({#2}) #3}
\def\ijmp#1#2#3{Int.~J.~Mod.~Phys.~{\bf A {#1}} ({#2}) #3}
\def\mpl#1#2#3{Mod.~Phys.~Lett.~{\bf A {#1}} ({#2}) #3}
\def\nc#1#2#3{Nuovo Cim.~{\bf {#1}} ({#2}) #3}
\def\ibid#1#2#3{{\it ibid.}~{\bf {#1}} ({#2}) #3}

\def\dd{\displaystyle}
\def\ud{\dd\frac{1}{\sqrt{2}}}
\def\ut{\dd\frac{1}{\sqrt{3}}}
\def\us{\dd\frac{1}{\sqrt{6}}}

\title{
%{\normalsize \mbox{ }\hfill
%\begin{minipage}{3cm}
%MPP-2005-118
%\end{minipage}}\\
%\vspace*{10mm}
%
\bf The problem of the initial
conditions in flavoured leptogenesis and
the tauon $N_2$-dominated scenario}
\author{{\large Enrico~Bertuzzo$^a$, Pasquale~Di Bari$^{b,c}$, Luca~Marzola$^{b}$}
\\
$^a$
{\it\small Scuola Normale Superiore and INFN,  Piazza dei cavalieri 7, 56126 Pisa, Italy}
\\
$^b$
{\it\small School of Physics and Astronomy},{\it\small University of Southampton,}
{\it\small  Southampton, SO17 1BJ, U.K.}\\
$^c$
{\it\small Department of Physics and Astronomy},{\it\small University of Sussex,}
{\it\small  Brighton, BN1 9QH, U.K.}
}

\maketitle \thispagestyle{empty}

\vspace{-10mm}
%\centerline{\date{\today}}

\begin{abstract}
We discuss the conditions to realize a scenario of `strong thermal leptogenesis',
where the final asymmetry is fully independent of the initial conditions,
taking into account both heavy and light neutrino flavor effects.
In particular, the contribution to the final asymmetry from an
initial pre-existing asymmetry has to be negligible.
We show that in the case of a hierarchical right-handed (RH)
neutrino mass spectrum, the only possible way is
a $N_2$-dominated leptogenesis scenario with a lightest RH neutrino mass
$M_1\ll 10^{9}\,{\rm GeV}$ and with a next-to-lightest RH neutrino
mass $10^{12}\,{\rm GeV}\gg M_2 \gg 10^{9}\,{\rm GeV}$.
This scenario necessarily requires the presence of a heaviest third RH neutrino specie.
Moreover, we show that the final asymmetry has to be dominantly produced in the tauon flavour
while the electron and the muon asymmetries have to be efficiently washed out by the lightest
RH neutrino inverse processes. Intriguingly, such seemingly special conditions
for successful strong thermal leptogenesis are naturally fulfilled
within $SO(10)$-inspired models. Besides the tauon $N_2$-dominated scenario,
successful strong thermal leptogenesis is also achieved
in scenarios with quasi-degenerate RH neutrino masses. We also
comment on the supersymmetric case. We also derive an
expression for the final asymmetry produced from leptogenesis taking fully
into account heavy neutrino flavour effects in the specific case
$M_1\gg 10^{12}\,{\rm GeV}$ (heavy flavoured scenario),
a result that can be extended to any other mass pattern.
\end{abstract}

\newpage

%%%%%%%%%%%%%%%%%%%%%%
\section{Introduction}
%%%%%%%%%%%%%%%%%%%%%%

With the discovery of neutrino masses and mixing
in neutrino oscillations experiments, leptogenesis \cite{fy},
a direct cosmological implication of the seesaw mechanism \cite{seesaw},
can be regarded as the most attractive model of baryogenesis.
When the minimal thermal scenario is considered, the predicted final asymmetry
depends on  18 additional parameters introduced by a minimal type I seesaw  with
three RH neutrinos.

The low energy neutrino experiments can only test the nine parameters in the light
neutrino mass matrix, whereas leptogenesis imposes just a condition on
the baryon-to-photon number ratio \cite{WMAP7}
\be\label{etaBobs}
\eta_B^{\rm CMB} = (6.2 \pm 0.15)\times 10^{-10} \, ,
\ee
so that testing the seesaw mechanism and leptogenesis, is seemingly impossible.

In order to circumvent this intrinsic obstacle, two complementary (but not mutually exclusive)
strategies are usually considered.  A first strategy  is to restrict the
parameter space imposing extra conditions coming from models of new physics.
A remarkable example is provided by $SO(10)$  models inspired conditions. Imposing
successful leptogenesis yields specific predictions on the
low energy neutrino parameters  \cite{lepgut,SO10}.
Recently, conditions required by leptogenesis within models of
discrete flavour symmetries attracted great attention as well \cite{jenkins,discrete}.
A second strategy is to look for additional  phenomenological
constraints on the seesaw mechanism parameter space in addition
to low energy neutrino experiments and leptogenesis.
Additional constraints within models beyond the SM
resulting for example from a determination of slepton masses in SUSY models,
from lepton flavor violation processes, from electric dipole moments,
from attempts to explain dark matter with RH neutrinos, from collider physics,
have been extensively studied  \cite{additionalpheno}.

However, even when the large number of free parameters is
restricted by some assumption or model of new physics or over-constrained by
supplementary phenomenologies, there is one further legitimate
conceptual issue to be addressed.
The predicted final asymmetry could  depend, in addition to the
seesaw parameters,  on the details of the cosmological history as well.
These could for example determine  the initial  abundance of the heavy right-handed neutrinos,
whose decays are responsible for the production of the asymmetry,
and a possible non-vanishing value of the $B-L$ asymmetry before
the onset of leptogenesis. In this respect it should be underlined that there
are many ways how a large pre-existing $B-L$
asymmetry could have been generated in the latest stage of inflation,
for example via  Affleck-Dine mechanism \cite{affleckdine},
by gravity  \cite{gravity}, or, after inflation and before the onset of leptogenesis,
even by more traditional GUT bosons decays \cite{GUTB}.

Apparently, this second obstacle can be more easily circumvented.
After all, standard Big Bang Nucleosynthesis provides already an example of
successful calculation of few observable cosmological quantities (the primordial nuclear
abundances) depending both on particle physics parameters and, potentially, on the details
of the history of the early Universe. By assuming  initial thermal conditions, for
a reheating temperature after inflation higher than $\sim 1\,{\rm MeV}$,
it is possible to calculate the primordial nuclear abundances
independently of the initial conditions.

Encouraged by this relevant example, one can assume a thermal scenario
of leptogenesis and calculate the final asymmetry almost independently
of  a detailed knowledge of the initial conditions and of many other possible
cosmological intervening complex processes.
The reheating temperature has then to be high enough that the Yukawa interactions alone
can efficiently thermalize the right-handed neutrino abundance. However, even though
this condition guarantees  a thermal production of the asymmetry,
it is still not sufficient to guarantee independence of the initial conditions.
Throughout the paper, we will refer
to a scenario of leptogenesis where the final asymmetry
is independent of the initial conditions, as {\em strong thermal leptogenesis}.
Moreover, we will always refer to the case of a hierarchical RH neutrino
spectrum, commenting in Section 5 on the case of scenarios with
quasi-degenerate RH neutrino masses.

A sufficient set of conditions for  strong thermal leptogenesis
usually involves additional constraints on the see-saw parameters
further restricting the parameter space.
More specifically, the independence of the initial conditions typically
translates into a condition of strong wash-out regime. In this case
any non-thermal or pre-existing contribution to the final asymmetry
is unavoidably washed-out by the same processes that are already predicted by the see-saw mechanism
(namely inverse processes involving RH neutrinos) without any need to impose new ad hoc
conditions or to extend the minimal seesaw framework.
The conditions for strong thermal leptogenesis are particularly simple when flavor effects are neglected.
They basically reduce to a simple condition on the value of the decay parameter $K_i$
of the RH neutrino $N_i$ responsible for the generation of the
asymmetry, $K_i \gtrsim K_{\star}\gg 1$ \cite{window}. The precise value of $K_{\star}$
depends just on the value of the initial pre-existing asymmetry. For example, as we will see,
for an initial very large pre-existing asymmetry ${\cal O}(1)$ one has  $K_{\star}\simeq 10$,
but it is in any case a constant value independent of the see-saw parameters.

On the other hand, when light neutrino flavor effects \cite{flavoreff} are taken into account,
the conditions become much more involved. An analysis of the conditions
for the independence of the initial RH neutrino abundance shows that the
constraints on the parameter space are even stronger \cite{flavorlep}.
They still result into a condition on the decay parameter $K_i\gtrsim K_{\star}$,
where, however, now $K_{\star}$  is not just a constant value but it depends
on the seesaw parameters. Therefore, the precise condition for the
independence of the initial conditions has to be determined
at each point in the seesaw parameter space.

 In \cite{nir} it was pointed out that
 there are no conditions able to guarantee an efficient wash-out of an arbitrary pre-existing asymmetry
 from  the lightest RH neutrino inverse processes,
unless these are lighter than $\sim 10^9\,{\rm GeV}$.
On the other hand, a complete study of the conditions for the wash-out of a pre-existing asymmetry,
accounting  also for the washout from the heavier RH neutrinos (heavy neutrino flavour effects)
and their compatibility with successful leptogenesis (an
important point in our analysis) is still missing.

One could simply think that conditions similar
to those found in \cite{flavorlep} for the independence of the RH neutrino abundance should hold.
However, in this paper, we will show that the conditions for
the wash-out of a pre-existing asymmetry are more elaborate and even more stringent.
They do not just involve the value of the decay parameters but even the RH neutrino mass
spectrum and the flavour composition of the final asymmetry themselves.
We will show, quite remarkably, that  when the mass spectrum is assumed to be hierarchical,
 only one specific mass pattern can allow for
 successful strong thermal leptogenesis
 \footnote{This conclusion is valid in non supersymmetric models. As we will comment
 in the last section, supersymmetric models offer an interesting loophole. Notice also
 that this conclusion does not translate into a prediction on the light neutrino mass spectrum,
 this would require more constrained models where the heavy neutrino and the light neutrino
 masses are linked.}.

Moreover we will show that when heavy neutrino flavors are taken into account,
because of new subtle effects, even the requirements for the
independence of the initial RH neutrino abundance become more stringent.
In the end we will show that the set of sufficient conditions
to be imposed in order to have an independence of the
initial asymmetries will be also sufficient for
an independence of the initial RH neutrino abundances.

In the second section we discuss how the wash-out of an initial asymmetry
can proceed in different ways depending on the specific RH neutrino
mass pattern and therefore we show how one has to distinguish different cases.
In the third section we start by studying the so called `heavy flavored scenario', where the leptons
can be treated as a coherent superposition of light ($e,\m,\t$) flavor eigenstates.
In the fourth section we consider `light flavored scenarios', concluding that
only a pattern where the lightest RH neutrino wash-out occurs in the three-light
flavored regime, for $M_1\ll 10^9\,{\rm GeV}$, can guarantee
a total wash-out of a pre-existing asymmetry. Furthermore we will show that if one also
imposes  successful leptogenesis, then necessarily one also has the
additional condition  $10^{12} \,{\rm GeV}\gg M_2 \gg 10^{9}\,{\rm GeV}$ and
that the final asymmetry has to be dominantly produced by the next-to-lightest RH neutrinos $N_2$
($N_2$-dominated scenario \cite{geometry}) in the tauon flavour.
In section 5 we make some final remarks (e.g. on supersymmetric models), we discuss some caveats,
and draw the conclusions.

%%%%%%%%%%%%%%%%%%%%%%%%%%%%%%%%%%%%%%%%%%%%
\section{Heavy and light neutrino flavors}
%%%%%%%%%%%%%%%%%%%%%%%%%%%%%%%%%%%%%%%%%%%%

The seesaw mechanism relies on the addition
to the SM lagrangian of RH neutrinos with Yukawa couplings and a Majorana mass term,
\begin{equation}\label{lagrangian}
\mathcal{L}= \mathcal{L}_{\rm SM} +i \overline{N_{R i}}\g_{\m}\partial^{\m} N_{Ri} -
h_{\a i} \overline{\ell_{L\a}} N_{R i} \tilde{\F} -
{1\over 2}\,M_i \overline{N_{R i}^c}N_{R i} +h.c.\quad (i=1,2,3,\quad \a=e,\m,\t) .
\end{equation}
For definiteness and simplicity we will consider the case of three RH neutrinos species,
though all data from low energy neutrino experiments can be explained within
a more minimal 2 RH neutrino model. Interestingly, we will see that the conditions of
strong thermal leptogenesis will require in the end the presence of
a third RH neutrino specie.

After spontaneous symmetry breaking, a Dirac mass term $m_D=v\,h$, is generated
by the vev $v=174$ GeV of the Higgs boson. In the see-saw limit, $M\gg m_D$,
the spectrum of neutrino mass eigenstates
splits in two sets: 3 very heavy neutrinos, $N_1,N_2$ and $N_3$
respectively with masses $M_1\leq M_2 \leq M_3$ and almost coinciding with
the eigenvalues of $M$, and 3 light neutrinos with masses $m_1\leq m_2\leq m_3$,
the eigenvalues of the light neutrino mass matrix
given by the see-saw formula \cite{seesaw},
\be
m_{\nu}= - m_D\,{1\over M}\,m_D^T \, .
\ee
Neutrino oscillation experiments measure two neutrino mass-squared
differences. For normal schemes one has
$m^{\,2}_3-m_2^{\,2}=\Delta m^2_{\rm atm}$ and
$m^{\,2}_2-m_1^{\,2}=\Delta m^2_{\rm sol}$,
whereas for inverted schemes one has
$m^{\,2}_3-m_2^{\,2}=\Delta m^2_{\rm sol}$
and $m^{\,2}_2-m_1^{\,2}=\Delta m^2_{\rm atm}$.
For $m_1\gg m_{\rm atm} \equiv
\sqrt{\Delta m^2_{\rm atm}+\Delta m^2_{\rm sol}}=
(0.050\pm 0.001)\,{\rm eV}$ \cite{gonzalez}
the spectrum is quasi-degenerate, while for
$m_1\ll m_{\rm sol}\equiv \sqrt{\D m^2_{\rm sol}}
=(0.0087\pm 0.0001)\,{\rm eV}$ \cite{gonzalez}
it is fully hierarchical (normal or inverted).
The most stringent upper bound on the
absolute neutrino mass scale is derived from
cosmological observations. Recently, quite a conservative
upper bound,
\be\label{bound}
m_1 < 0.19\,{\rm eV} \, \hspace{5mm} (95\%\, {\rm CL}) \, ,
\ee
has been obtained by the
WMAP collaboration combining CMB, baryon acoustic oscillations
and the Hubble Space Telescope measurement of $H_0$  \cite{WMAP7}.

With leptogenesis, this simple extension
of the Standard Model is also able to explain
the observed baryon asymmetry of the Universe eq.~(\ref{etaBobs}).
This is generated by the
$C\!P$ violating decays of the RH neutrinos
into leptons, $N_i \rightarrow {\ell}_i + H^{\dagger}$,
and into anti-leptons, $N_i \rightarrow \bar{\ell}_i + H$,
producing a lepton asymmetry
that for temperatures $T\gtrsim 100\,{\rm GeV}$ is partially converted into
a baryon asymmetry by sphaleron ($B-L$ conserving) processes.
The observed baryon asymmetry can then be calculated from the final $B-L$
asymmetry simply using
\be\label{etaB}
\eta_B=a_{\rm sph}\,{N_{B-L}^{\rm f}\over N_{\gamma}^{\rm rec}}
\simeq 0.96\times 10^{-2}\,N_{B-L}^{\rm f} \, ,
\ee
where we indicate with $N_X$
any particle number or asymmetry $X$ calculated in a portion
of co-moving volume containing one heavy neutrino in ultra-relativistic
thermal equilibrium, i.e. such that $N^{\rm eq}_{N_2}(T\gg M_2)=1$.

The leptons produced in  $N_i$ decays can be described in terms
of quantum states that we indicate with $|{\ell}_i\rangle$.
They have a flavor composition given by
\be
|{\ell}_i\rangle = \sum_{\a}\,{\cal C}_{i\a}\,|{\ell}_\a \rangle \, , \;\;\;\;
{\cal C}_{i\a} \equiv  \langle {\ell}_\a|\ell_i \rangle  \;\;\;\; (\a=e,\mu,\t) \,
\ee
that is in general distinct for each heavy neutrino flavour $i$. We will refer  to them
as `heavy neutrino flavours' lepton quantum states and analogously we will refer to the
$|{\ell}_{\alpha}\rangle$ as the `light neutrino flavours' lepton quantum states.
Similarly,  one can write for the anti-leptons
\be
|\bar{{\ell}_i'}\rangle = \sum_{\a}\,\bar{{\cal C}}_{i\a}\,|\bar{{\ell}}_\a \rangle \, , \;\;
\bar{{\cal C}}_{i\a} \equiv  \langle \bar{{\ell}}_\a|\bar{\ell}_i' \rangle \, , \;\; (\a=e,\mu,\t) \, .
\ee
Notice that while the light neutrino flavours quantum states form an orthonormal basis,
\be
\langle {\ell}_\a|{\ell}_\b \rangle = \d_{\a\b} \, \;\;\;\; \mbox{\rm and} \;\;\;\;
 \langle \bar{{\ell}}_\a|\bar{{\ell}}_\b \rangle = \d_{\a\b} \, ,
\ee
in general the heavy neutrino flavour quantum states do not \cite{nardi1}, since in general
the quantities $\langle \ell_j|\ell_{i}\rangle$ do not vanish. In the
Appendix we give an expression for the $p_{ij}\equiv |\langle \ell_j|\ell_{i}\rangle|^2$
in terms of the Dirac mass matrix  showing this point.
Analogously, the pre-existing lepton quantum states $|{\ell}_{\rm p}\rangle$ and anti-lepton quantum states
$|\bar{\ell}^{\,'}_{\rm p}\rangle$ also have a flavour composition, given respectively by
\be\label{flavcomprlep}
|{\ell}^{\rm p}\rangle = \sum_{\a}\,{\cal C}_{p\a}\,|{\ell}_\a \rangle \, , \;\;\;\;
{\cal C}_{p\a} \equiv  \langle {\ell}_\a|\ell^{\rm p} \rangle  \;\;\;\; (\a=e,\mu,\t)
\ee
and
\be\label{flavcomprantilep}
|\bar{\ell }^{{\rm p}\,'} \rangle = \sum_{\a}\,\bar{\cal C}_{p\a}\,|\bar{\ell}_\a \rangle \, , \;\;\;\;
\bar{\cal C}_{p\a} \equiv  \langle \bar{\ell}_\a|\bar{\ell}^{\rm p\,'} \rangle  \;\;\;\; (\a=e,\mu,\t) \, .
\ee
These leptons are those ones that, in the standard cosmological picture, would  be produced
at the end or soon after inflation, before the RH neutrino processes become effective.

In general the flavour composition of leptons and of  $C\!P$ conjugated anti-leptons
are different. However, we will first derive our conclusions assuming that
they are equal, i.e.  ${\cal C}_{p\a}=\bar{\cal C}_{p\a}$ and then,
in section 5, we will point out that even allowing for a
different flavour composition of the pre-existing leptons and anti-leptons
our conclusions do not change.

While flavor blind gauge interactions preserve  the coherence of the lepton quantum states,
Yukawa charged lepton interaction are flavor sensitive and act as a potential source of decoherence
\cite{nardi1,abada1} which can be neglected only for sufficiently high values of the decaying
RH neutrino masses \cite{zeno}
\be\label{cond1}
M_i \gg 10^{12}\,{\rm GeV} \, .
\ee
On the other hand, if
\be\label{cond2}
10^{12}\,{\rm GeV}  \gg M_i \gg 10^{9}\,{\rm GeV}  \, ,
\ee
the tauon charged lepton interactions are on average fast enough to
destroy the  coherent evolution of the quantum superposition
of the tauon and of the its orthogonal component (superposition
of the electron and of the muon component) before the lepton quantum state
interacts inversely with an Higgs boson to produce a RH neutrino $N_i$.
Therefore, in this two light flavor regime,
the lepton quantum states inverse decay as  an incoherent mixture of
tauon flavor eigenstates plus the (still) coherent $\tau$ orthogonal
superposition of the electron and of the muon components. Therefore, the wash-out
from inverse decays has to be taken into account separately on the
asymmetry in the tauon flavour and on the asymmetry in the tauon orthogonal flavour
component that we will indicate with $\tilde{\tau}$
\footnote{In general, throughout the paper, given a flavour $x$, we will indicate  with $\tilde{x}$
the set of the orthogonal flavours corresponding to the plane orthogonal to $x$ in flavour space.}.

Finally, if
\be\label{cond3}
M_i \ll 10^{9}\,{\rm GeV} \, ,
\ee
then even the coherence of the superposition  of the  electron and of the muon component
breaks down and the lepton quantum states, at $T\sim M_i$, have to be described as
a fully  incoherent
mixture of three light flavor eigenstates (three-flavor regime).

In our discussion we will assume that the
three RH neutrino masses satisfy one of the previous conditions
(eq.~(\ref{cond1}), eq.~(\ref{cond2}) or eq.~(\ref{cond3})),
in such a way that the collisions of the lepton doublets with the charged leptons
are fast enough that a classical Boltzmann description provides
a good approximation \cite{densitymatrix,dr,fv}. We will also assume a hierarchical
RH neutrino mass spectrum with $M_{i+1}\gtrsim 3\,M_i$ in a way that decays and wash-out
for each RH neutrino specie occur in separate stages that do ont overlap with each other \cite{beyond}.

There is another source of  decoherence of the leptonic quantum states $|{\ell}_i\rangle$
produced from $N_i$ RH neutrino decays and on the the $|{\ell}^{\rm p}\rangle$.
Notice that in general the $|{\ell}_i\rangle$  do not form an orthonormal basis (see Appendix),
meaning that in general $\langle {\ell}_i |{\ell}_{j\neq i}\rangle \neq 0$.
At the same time the $|{\ell}^{\rm p}\rangle$ have an
arbitrary flavour composition (e.g. an equal mixture of electron, muon
and tauon components). This implies that  the
quantum lepton states $|{\ell}_i\rangle$ and $|{\ell}_{\rm p}\rangle$, after an inverse
process producing a RH neutrino $N_{j\neq i}$, become an incoherent mixture of
a $|{\ell}_j\rangle$ component and of a $|{\ell}_j\rangle$  orthogonal component
that, in our convention, it will be indicated by $|{\ell}_{\tilde{j}_i}\rangle$ \cite{bcst,nir}.
Also in this case, we will employ a classical Boltzmann description for the evolution
of the asymmetries, assuming that
at each inverse process, the $|{\ell}_i\rangle$  (or the $|{\ell}_{\rm p}\rangle$) quantum state
instantaneously collapses either into the $|{\ell}_j\rangle$ parallel component
or into its orthogonal one that we will indicate with $|{\ell}_{\tilde{j}_i}\rangle$
(or $|{\ell}_{\tilde{j}_{\rm p}}\rangle$). Again the classical picture holds when
the interactions are fast enough \cite{densitymatrix,dr,fv}, in this case with respect to the expansion
rate since there are no other involved active processes
\footnote{Within a full quantum density matrix formalism, corrections to the classical
picture are expected to be maximal in an intermediate
regime of mild interaction rates. However, they cannot be particularly relevant in the case
of hierarchical RH neutrino mass patterns. This is because the $N_j$ inverse processes occur when the
the $|{\ell}_i\rangle$ production from the $N_i$ decays has already switched off and
therefore, there is no overlap between the $N_i$ decays and the $N_j$ inverse decays.
In this situation the damping of the density matrix off-diagonal terms is rather fast.
As we will see our conclusions will not rely on a detailed kinetic description.
On the other hand a density matrix treatment seems to be unavoidable if one considers
RH neutrino mass patterns beyond the  hierarchical limit, when decays and inverse processes
involving different RH neutrino species occur simultaneously and compete with each other.}.

The potential simultaneous interplay of six different flavors of leptons,
the three heavy ones and three light ones,
the non-orthogonality of the three heavy flavors in contrast with the orthogonality
of the three light flavors, yield different scenarios corresponding to different
RH neutrino mass patterns that need to be discussed separately.
In the next sections we will discuss each case separately but
it is useful to make before some general comments.

Because of the linearity of the Boltzmann equations,
the final $B-L$ asymmetry is the sum of two contributions,
\be
N_{B-L}^{\rm f}=N_{B-L}^{\rm p,f}+ N_{B-L}^{\rm lep,f} \, .
\ee
The first term is the residual value of a pre-existing asymmetry,
after the RH neutrinos wash-out, whereas the second term is the final value of the
$B-L$ asymmetry produced by the RH neutrino decays, the product of
leptogenesis genuinely depending only on the seesaw parameters.

 The main goal of this paper is to show under which conditions one can have simultaneously
 both successful leptogenesis and a negligible contribution from the pre-existing
 asymmetry, i.e.
 \be\label{primarycond}
 |N_{B-L}^{\rm p,f}| \ll |N_{B-L}^{\rm lep,f}| \, .
 \ee
 Only in this way it is guaranteed that the successful leptogenesis condition,
 $\eta_B^{\rm f} =\eta_B^{CMB}$, really constraints the seesaw parameter space.
 Notice that we cannot completely ignore  $N_{B-L}^{\rm lep,f}$,
  since, after having found the conditions for (\ref{primarycond}) to hold,
  we have also to check that these do not prevent successful leptogenesis.
 As we will see, this check will not require an explicit  calculation of $N_{B-L}^{\rm lep,f}$
 and  we will mainly focus on the evolution of $N_{B-L}^{\rm p,f}$.
 However, in the Appendix, we also explicitly derive the solution for
 $N_{B-L}^{\rm lep,f}$ for a specific RH neutrino mass pattern,
 in the so called `heavy flavoured scenario'.

Notice that we will  neglect throughout sections 3 and 4
a new effect studied in \cite{flcoup}, so called light neutrino flavour coupling.  However,
we will comment in section 5 about the kind of impact that
light neutrino flavour coupling might have on our conclusions.

It is also useful to make a general remark about the flavour composition of the pre-existing
leptons and anti-leptons.
The asymmetry $N_{B-L}^{\rm p,i}$ is distributed not only in the lepton doublets
but also in the RH charged leptons and in the quarks.
We will always assume that $T_i\lesssim 10^{14}\,{\rm GeV}$
so that sphaleron processes are in equilibrium and the asymmetry distribution
among the different particle species can be calculated from equilibrium conditions \cite{sphalerons}.
In particular the asymmetry in the lepton doublets is related to the $B-L$ asymmetry by
\be
N_{L}= \alpha_L \, N_{B-L} \, ,
\ee
with $\a_L\simeq -2/3$. Assuming that only the RH neutrino decays
and inverse processes can modify the $B-L$ asymmetry,
any change of $N_{B-L}$ can be only triggered by
a modification of $N_L$, given by the difference
of the number of leptons and anti-leptons.

On practical grounds, our objective is to calculate
the final value of the wash-out factor of a pre-existing asymmetry defined as
\be
w^{\rm f} \equiv {N_{B-L}^{\rm p,f}\over N_{B-L}^{\rm p,i}} \, .
\ee
Conservatively, for a pre-existing asymmetry $N_{B-L}^{\rm p,i}={\cal O}(1)$, e.g. generated
by an Affleck-Dine mechanism during inflation \cite{affleckdine}, the independence of the initial conditions
requires $w^{\rm f}\ll 10^{-8}$.

So far, this wash-out factor has been calculated only
in a completely unflavoured case \cite{window}, i.e. neglecting both light and heavy neutrino flavour effects.
It is useful, also in order to set up the notation, to review briefly this simple calculation.

In this simple picture, the pre-existing asymmetry has no flavour composition.
Assuming  $M_3, M_2 \gg T_i \gg M_1$,  one
can just consider the wash-out from the lightest RH neutrinos $N_1$.
 The $B-L$ asymmetry is then described by a particularly simple set
 of just two Boltzmann equations \cite{pedestrians},
\begin{eqnarray}\label{noflke}
{dN_{N_1}\over dz_1} & = &
-D_1\,(N_{N_1}-N_{N_1}^{\rm eq}) \;,
\hspace{25mm}  \label{dlg1} \\\label{unflke}
{dN_{B-L}\over dz_1} & = &
\varepsilon_1\,D_1\,(N_{N_1}-N_{N_1}^{\rm eq})-
N_{B-L}\,W_1(z_1)  \, ,
\label{dlg2}
\end{eqnarray}
where we defined $z_i \equiv M_i/T$ ($i=1,2,3$). The expansion rate is expressed as
\begin{equation}
H= \sqrt{8\,\pi^3\,g_{\star}\over 90} {M_i^2\over M_{\rm Pl}}\,{1\over z_i^{2}}
\simeq 1.66\,\sqrt{g_{\star}}\,{M_i^2\over M_{\rm Pl}}\,{1\over z_i^{2}} \, ,
\end{equation}
having indicated with $g_{\star}=g_{SM}=106.75$ the total number of degrees of freedom
and with $M_{\rm Pl}=1.22\,\times\, 10^{19}\,{\rm GeV}$ the Planck mass.

If we indicate with $\Gamma_i$ the decay rate of the RH neutrinos $N_i$ into leptons,
and with $\bar{\Gamma}_i$ the decay rate into anti-leptons,
the total decay rate, $\G_{{\rm D},i} \equiv \G_i+\bar{\G}_i =
\widetilde{\Gamma}_i\, \langle 1/\gamma\rangle$, is the product of the decay
width times the thermally averaged dilation factor,
that can be expressed in terms of the ratio of the modified Bessel functions,
$\langle 1/\gamma\rangle={\cal K}_1(z)/ {\cal K}_2(z)$.

The decay parameters are then defined as
$ K_i\equiv \widetilde{\Gamma}_i / H(z_i=1) = \widetilde{m}_i / m_\star$,
and can be expressed in terms the effective neutrino masses \cite{plumacher}
and of the equilibrium neutrino mass \cite{orloff,window},  respectively given by
\be
\widetilde{m}_i=v^2\frac{(m^{\dagger}_D\,m_D)_{ii}}{M_i}  \hspace{5mm}
\mbox{\rm and} \hspace{5mm}
m_\star=\frac{16 \pi^{5/2} \sqrt{g_\star}}{3 \sqrt{5}}\frac{v^2}{M_{Pl}}\simeq 1.08
\times 10^{-3}~\mathrm{eV} \, .
\ee
The decay terms and the related wash-out terms are given by
\be\label{WID}
D_i \equiv {\G_{{\rm D},i}\over H\,z}=K_i\,z_i\,
\left\langle {1\over\gamma_i} \right\rangle \, ,
\hspace{5mm}  \hspace{5mm}
W_i(z_i) = {1\over 2}\,D_i(z_i)\,N_{N_i}^{\rm eq}(z_i)
={1\over 4}\,K_i\,{\cal K}_1(z_i)\,z_i^3 \, ,
\ee
where the equilibrium abundance  has been also expressed
in terms of the modified Bessel functions.
Defining the total $C\!P$ asymmetries as
\be
\ve_i \equiv -{\Gamma_i-\bar{\Gamma}_i\over \Gamma_i+\bar{\Gamma}_i} \, ,
\ee
the final asymmetry is then simply given by
\be
N_{B-L}^{\rm f} = N_{B-L}^{\rm p,i}\,e^{-{3\pi\over 8}\,K_1} + \ve_1\,\k(K_1) \, ,
\ee
where $\k(K_1)$ is the efficiency factor. Therefore, in this case
one has $N_{B-L}^{\rm lep}=\ve_1\,\k(K_1) $ while the wash-out factor of the
pre-existing asymmetry is simply given by $w^{\rm f}= e^{-3\pi K_1/8}$.

It is then sufficient to impose $K_1\gtrsim 10$ in order to have $w^{\rm f} \lesssim 10^{-8}$,
ensuring the wash-out of a ${\cal O}(1)$ pre-existing asymmetry
\footnote{For definiteness, throughout the paper we will also
refer to the wash-out of a ${\cal O}(1)$ pre-existing asymmetry.}.
Notice moreover that if $T_i\gtrsim M_2$, then any  asymmetry
produced from the next-to-lightest RH neutrinos is washed out by the $N_1$
inverse processes in the same way as the pre-existing asymmetry.

When flavour effects are taken into account, these conclusions change  drastically.
In \cite{nir} it was shown that a simple condition $K_1\gtrsim 10$ is not sufficient to
guarantee the complete wash-out of a ${\cal O}(1)$ pre-existing asymmetry.
This is possible only if
$M_1\ll 10^{9}\,{\rm GeV}$ and if $K_{1\alpha}\equiv p_{1\alpha}\,K_1 \gtrsim 10$
for all $\alpha=e,\mu,\tau$,
where we defined $p_{i\alpha}\equiv |\langle {\ell}_{\alpha}|{\ell}_i \rangle|^2$.
However, such a drastic condition is not compatible with successful leptogenesis
since  any asymmetry produced from the heavier RH neutrinos
is washed-out together with the pre-existing asymmetry as well while
the lightest RH neutrino $C\!P$ asymmetries are too low
for an asymmetry to be generated from $N_1$ decays after the freeze out of
wash-out processes.

We have therefore to extend the analysis to a more general case where
the assumption $M_3, M_2 \gg T_i \gg M_1$ is relaxed,
pinning down the conditions for successful strong thermal leptogenesis.
The discussion has to be specialized considering all possible mass patterns
characterized by a different interplay between light and heavy neutrino flavour effects.
We start discussing the so called `heavy flavored scenario',
where all the three RH neutrino masses satisfy the condition eq.~(\ref{cond1}).

%%%%%%%%%%%%%%%%%%%%%%%%%%%%%%%%%
\section{Heavy flavored scenario}
%%%%%%%%%%%%%%%%%%%%%%%%%%%%%%%%%

In this scenario all three RH neutrino
masses are much heavier than about $10^{12}\,{\rm GeV}$ (see Fig. 1).
In this case the $|{\ell}_i\rangle$'s evolve coherently since the
charged lepton interactions are ineffective in measuring their
light neutrino flavor composition.
\begin{figure}[htbp]
\begin{center}
\includegraphics[height=5cm,width=5cm]{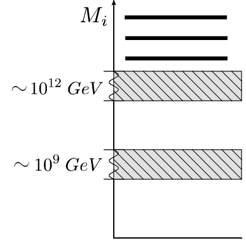}
\end{center}
\caption{The mass pattern corresponding to the heavy flavoured scenario.}
\label{fig1}
\end{figure}
We also assume $T_i\gg M_3$ so that all three RH neutrinos can
wash-out the pre-existing asymmetry. If we prove that
some fraction of the pre-existing asymmetry survives in this most conservative case,
then  some (at least equal) fraction necessarily  survives
if $T_i \ll M_3$ as well.

\subsection{First stage: $T_i > T \gg M_3$}

There are different stages in the evolution of $N_{B-L}^{\rm p}(z)$.
In a first stage, for $T_i > T \gg M_3$, all RH neutrino
processes are ineffective  and the $B-L$ asymmetry remains constant.
The flavour composition of the pre-existing lepton doublets quantum states can be
regarded as a coherent superposition of a ${\ell}_3$ parallel component
and of a  ${\ell}_3$ orthogonal component, explicitly
\footnote{Notice that with the notation ${\ell}_{\tilde{3}}$ we mean
the projection of $|{\ell}^{\rm p}\rangle$ on the plane orthogonal to $|{\ell}_3\rangle$
so that $\langle {\ell}_3|{\ell}_{\tilde{3}} \rangle = 0$.
More precisely we should write ${\ell}_{\tilde{3}_{\rm p}}$ but we imply
the subscript ${\rm p}$ in order to simplify the notation.
We will do the same for the projections on the planes orthogonal to
$|{\ell}_2\rangle$  and $|{\ell}_1\rangle$.}
\be
|{\ell}^p\rangle = {\cal C}_{p3} |{\ell}_3 \rangle + {\cal C}_{p\tilde{3}}
|{\ell}^{\rm p}_{\tilde{3}} \rangle \, ,
\;\;\; \mbox{\rm with} \;\;\; p_{p3} + p_{p\tilde{3}}=1 \, ,
\ee
where $p_{p3} \equiv |{\cal C}_{p3}|^2$ and $p_{p\tilde{3}} \equiv |{\cal C}_{p\tilde{3}}|^2$.
This decomposition is pictorially represented in the upper-right panel of Fig. 2.
\begin{figure}[htbp]
\begin{center}
\subfigure[$T \gg M_3$]{\includegraphics[width=7.2cm]{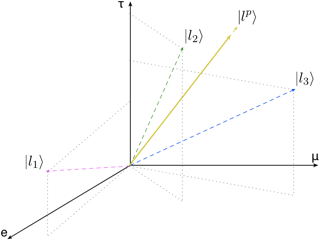}}
\subfigure[$T\sim M_3$]{\includegraphics[width=7.2cm]{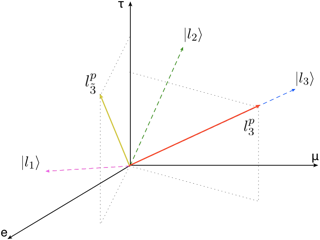}}\\
\subfigure[$T\sim M_2$]{\includegraphics[width=7.2cm]{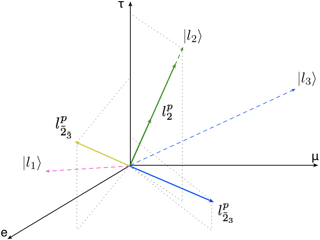}}
\subfigure[$T\sim M_1$]{\includegraphics[width=7.2cm]{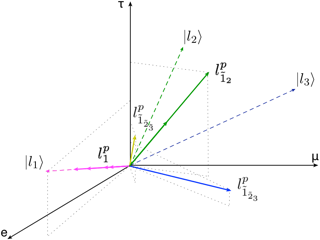}}\\
\end{center}
\caption{The four stages of the heavy flavoured scenario. In a first stage at $T\gg M_3$ (a) the
pre-existing leptons are a coherent superposition of light flavor eigenstates.
At $T\sim M_3$ (b) the $N_3$ decays and inverse processes break the coherent evolution of
$|{\ell}_{\rm p}\rangle$ that becomes an incoherent mixture of a $|{\ell_3}\rangle$ component and of a $|{\ell}_{\tilde{3}}\rangle$ component.
At $T\sim M_2$ (c), the $|{\ell}_3\rangle$ and the  $|{\ell}_{\tilde{3}}\rangle$ components are
both re-projected on a $|{\ell}_2\rangle$ component and on a $|{\ell}_{\tilde{2}}\rangle$ component.
We indicate with ${\ell}^{\rm p}_i$ or ${\ell}^{\rm p}_{\tilde{i}_j}$,
those components of pre-existing leptons with a certain flavour composition, $i$ and $\tilde{i}_j$,
that give a  contribution to the residual  pre-existing asymmetry and that experienced a
different wash-out history. For example ${\ell}^{\rm p}_i$ is the component of ${\ell}^{\rm p}$
that experienced only a wash-out from $N_i$ RH neutrinos. On the other hand for example
${\ell}^{\rm p}_{\tilde{1}_2}$ is the component of ${\ell}^{\rm p}$ that has been
first projected along ${\ell}_2$, undergoing the wash-out from the $N_2$'s,
and then along a direction orthogonal to ${\ell}_1$, escaping the wash-out from the $N_1$'s.
Notice that at a given stage components with a different wash-out history
might end up in the same quantum state.
At each stage the red components are those undergoing wash-out
while the yellow ones are those that are unwashed.
At $T\sim M_1$ (d), the $|{\ell}_2\rangle$ and the $|{\ell}_{\tilde{2}}\rangle$ components
are finally projected on a $|{\ell}_1 \rangle$ and on a $|{\ell}_{\tilde{1}}\rangle$ component.
Again the different arrows just simply track the different histories. At this stage
there are 8 contributions to the final asymmetry that experienced a different wash-out.
The yellow component is  completely unwashed.}
\label{fig1}
\end{figure}
The same decomposition can be made for the pre-existing anti-leptons and under the given assumptions
one has $\bar{\cal C}_{p3}={\cal C}_{p3}$ and
$\bar{\cal C}_{p\tilde{3}}={\cal C}_{p\tilde{3}}$.
Correspondingly the pre-existing $B-L$ asymmetry can be also decomposed as
\be\label{decomposition}
N_{B-L}^{\rm p,i}=N_{\Delta_3}^{\rm p,i} + N_{\Delta_{\tilde{3}}}^{\rm p,i} \, ,
\ee
where we defined $N_{\Delta_3}^{\rm p,i}=p_{p3}\,N_{B-L}^{\rm p,i}$ and
$N_{\Delta_{\tilde{3}}}^{\rm p,i}=(1-p_{p3})\,N_{B-L}^{\rm p,i}$.

\subsection{Second stage: $M_3\gtrsim T \gtrsim T_{B3}$}

Let us now discuss a second stage for $M_3\gtrsim T \gtrsim T_{B3}$,
where $T_{B3}\simeq M_3/z_{B3}$ is the
freeze-out temperature of the $N_3$ inverse processes and
$z_{B3}\simeq 2+4\,K_{3}^{0.13}\,e^{-{2.5/K_{3}}}={\cal O}(1-10)$  \cite{beyond}.
An interaction of a quantum lepton state $|{\ell}^p\rangle$ with a Higgs boson
can be regarded, in a classical statistical picture, as a measurement process where
there is a probability $p_{p3}$ that $|{\ell}^p\rangle$ is measured
as a ${\ell}_3$ producing a $N_3$ in the inverse decay and a probability $1-p_{p3}$ that is measured
as a ${\ell}_{\tilde{3}}$ and in this  case no inverse process occurs
\footnote{This is  analogous to what happens in active-sterile neutrino
oscillations described in terms classical Boltzmann equations \cite{activesterile,fv}, where
 the orthogonal component here plays the role of the  sterile component.}.
In this way only the component $N_{\Delta_3}^{\rm p,i}$  of the $B-L$ asymmetry
is washed out, while the orthogonal  component $N_{\Delta_{\tilde{3}}}^{\rm p,i}$ is basically untouched
(see again the upper right panel in Figure 2)
\footnote{Notice that we are neglecting what can be called
`heavy neurino flavour coupling' \cite{nir} due to spectator processes,
mainly to the Higgs asymmetry. When this is
taken into account the evolution of the two components $N_{\Delta_3}^{\rm p,i}$ and
$N_{\Delta_{\tilde{3}}}^{\rm p,i}$ is not completely independent of each other
but couple to some extent. The result is that $N_{\Delta_3}^{\rm p,i}$
is not completely washed-out while the component $N_{\Delta_{\tilde{3}}}^{\rm p,i}$
partly is. However, an account of this effect does not change our conclusions,
as we point out in Section 5.}.

We also have to take into account the production of the  ${\ell}_3$'s and of the
$\bar{\ell}'_3$'s from the $C\!P$ violating decays of the heaviest RH neutrinos $N_3$.
Within the adopted classical statistical picture, an interaction acts as a
measurement process on the
quantum state that collapses in one of the two components. Therefore,
we can use  Boltzmann equations to describe the evolution of the asymmetry.
These   can be simply written as
\begin{eqnarray}
{dN_{N_3}\over dz_3} & = & -D_3\,(N_{N_3}-N_{N_3}^{\rm eq}) \;, \hspace{25mm}   \\\label{heavyflke}
{dN_{\D_3}\over dz_3} & = & \ve_3\,D_3\,(N_{N_3}-N_{N_3}^{\rm eq})-W_3\,N_{\Delta_3} \, , \\
{dN_{\D_{\tilde{3}}}\over dz_3} & = & 0 \, .
\end{eqnarray}
As we said, we give in the Appendix the calculation of the asymmetry produced from the decays,
focusing here only on the evolution of the residual pre-existing asymmetry.
The residual pre-existing asymmetry  after this stage, at $T\sim T_{B3}$, will be  given then by
\bea
N_{B-L}^{\rm p}(T_{B3}) & = & N_{\Delta_3}^{\rm p,i}\,e^{-{3\pi\over 8}\,K_3} + N_{\Delta_{\tilde{3}}}^{\rm p,i} \\
& = & p_{p3}\,N_{B-L}^{\rm p,i}\,e^{-{3\pi\over 8}\,K_3} + (1-p_{p3})\,N_{B-L}^{\rm p,i} \, ,
\eea
corresponding to a wash-out factor
\be
w(T_{B3})=p_{p3}\,e^{-{3\pi\over 8}\,K_3} + 1-p_{p3} \, .
\ee
As one can see, there is no condition that can be imposed on the see-saw parameters able to
guarantee an efficient wash-out of a pre-existing asymmetry with a generic
flavour composition ($p_{p3}\neq 1$). If we impose $K_3\gg 1$,
at the end of the stage at $T\sim T_{B3}$, the lepton doublets are an incoherent mixture of
${\ell}_3$ and ${\ell}_{\tilde{3}}$ and analogously the anti-leptons are an incoherent mixture of
$\bar{\ell}_3$ and $\bar{\ell}_{\tilde{3}}$. The asymmetry
in the ${\ell}_3$ component is efficiently washed out (the first term in the RH side
of the previous equation) but the asymmetry in the orthogonal component survives.

\subsection{Third stage: $T_{B3}\gtrsim T \gtrsim T_{B2}\sim M_2/z_{B2}$}

Let us then consider  the subsequent third stage, for $T_{B3}\gtrsim T \gtrsim T_{B2}\sim M_2/z_{B2}$.
At $T\sim M_2$ the $N_2$ inverse processes start to collapse the
$|{\ell}_3\rangle$ and the $|{\ell}_{\tilde{3}}\rangle$ either as a $|{\ell_2}\rangle$
or as the orthogonal component $|{\ell}_{\tilde{2}}\rangle$.
We can therefore repeat exactly the same decomposition as in the previous stage and write
the residual pre-existing asymmetry at the end of the previous stage as the sum
of two terms,
\be
N_{B-L}^{\rm p}(T_{B3}) = N_{\Delta_2}^{\rm p}(T_{B3}) + N_{\Delta_{\tilde{2}}}^{\rm p}(T_{B3}) \, ,
\ee
where
\bea
N_{\Delta_2}^{\rm p}(T_{B3})& = & p_{32}\,N_{\Delta_3}^{\rm p}(T_{B3})+
                         p_{\tilde{3}2}\,N_{\Delta_{\tilde{3}}}^{\rm p}(T_{B3}) \\
                        & = & p_{32}\,p_{p3}\,N_{B-L}^{\rm p,i}\,e^{-{3\pi\over 8}\,K_3} +
                              p_{\tilde{3}2}\,(1-p_{p3})\,N_{B-L}^{\rm p,i} \, ,
\eea
and
\bea\label{ND2TB3}
N_{\Delta_{\tilde{2}}}^{\rm p}(T_{B3})& = & (1-p_{32})\,N_{\Delta_3}^{\rm p}(T_{B3})+
                                    (1-p_{\tilde{3}2})\,N_{\Delta_{\tilde{3}}}^{\rm p}(T_{B3}) \\
                        & = & (1-p_{32})\,p_{p3}\,N_{B-L}^{\rm p,i}\,e^{-{3\pi\over 8}\,K_3} +
                              (1-p_{\tilde{3}2})\,(1-p_{p3})\,N_{B-L}^{\rm p,i} \, .
\eea
These are the two terms of the asymmetry  that have to be used as initial condition
at the beginning of the $N_2$ wash-out stage.
The Boltzmann equations are obtained from the eqs. (\ref{heavyflke}) written
in the previous stage with the simple replacement of the label $3\rightarrow 2$,
explicitly
\begin{eqnarray}
{dN_{N_2}\over dz_2} & = & -D_2\,(N_{N_2}-N_{N_2}^{\rm eq}) \;, \hspace{25mm}   \\\label{heavyflke}
{dN_{\D_2}\over dz_2} & = & \ve_2\,D_2\,(N_{N_2}-N_{N_2}^{\rm eq})-W_2\,N_{\Delta_2} \, , \\
{dN_{\D_{\tilde{2}}}\over dz_2} & = & 0 \, .
\end{eqnarray}
We can therefore straightforwardly express the residual
pre-existing asymmetry at $T\sim T_{B2}$ as
\footnote{Notice that the notation ${\ell}^{\rm p}_{\tilde{x}_{\tilde{y}}}$ indicates
a lepton with  flavour obtained by projecting the pre-existing
quantum lepton state $|{\ell}_{\rm p}\rangle$ first on a plane orthogonal to the flavour $y$ and then
to the plane orthogonal to the flavour $x$. Correspondingly
$N^{\rm p}_{\Delta_{\tilde{x}_{\tilde{y}}}}$ indicates the asymmetry in that flavour.}
\be\label{NBmLTB2}
N_{B-L}^{\rm p}(T_{B2}) = N_{\Delta_2}^{\rm p}(T_{B3})\,e^{-{3\pi\over 8}\,K_2}+
                      N_{\Delta_{\tilde{2}_{\tilde{3}}}}^{\rm p}(T_{B3}) \, .
\ee
Imposing $K_2,K_3\gtrsim 10$, one can neglect
all terms exponentially suppressed so that the
wash-out factor reduces to
\be
w(T_{B2})\simeq (1-p_{\tilde{3}2})\,(1-p_{p3}) \, .
\ee
This result shows that in general, even at this stage
after the wash-out both from $N_3$ and from $N_2$ inverse processes,
there is no condition that one can impose on the see-saw parameters able to guarantee
an efficient wash-out of a generic pre-existing asymmetry.

\subsection{Fourth stage: $T_{B2}\gtrsim T \gtrsim T_{B1}\sim M_1/z_{B1}$}

The  wash-out from the lightest RH neutrinos
can now be straightforwardly calculated going along
the same lines as in the previous stages.
At the end of the $N_1$ wash-out, at $T\sim T_{B1}$, the asymmetry
freezes at its final value and one has
\be\label{NBmLpfhf}
N_{B-L}^{\rm p}(T_{B1})= N_{B-L}^{\rm p,f} = N_{\Delta_1}^{\rm p}(T_{B2})\,e^{-{3\pi\over 8}\,K_1}+
                      N_{\Delta_{\tilde{1}}}^{\rm p}(T_{B1}) \, .
\ee
The final residual value of the pre-existing asymmetry is the sum of 8 terms (they are explicitly written
in the Appendix), where 7 of them
undergo the wash-out exponential suppression either of just one (three terms), or of two (three terms)
or of all three (1 term) RH neutrinos.  There is clearly one component that
escapes the wash-out of all three RH neutrinos. Imposing $K_1,K_2,K_3\gtrsim 10$,
all contributions that undergo the wash-out of at least one RH neutrino are
suppressed below the observed value. The final wash-out factor will be
then dominated only by the contribution coming from the
completely unwashed term and it is therefore given by
\be\label{wfhf}
w^{\rm f} \simeq (1-p_{\tilde{2}_{\tilde{3}}1})\,(1-p_{\tilde{3}2})\,(1-p_{p3}) \, .
\ee
It is clear that, barring very special situations, that should be realized with great
precision, the wash-out of a pre-existing large asymmetry cannot be enforced.
These special situations are realized either when the pre-existing
leptons coincide with the  ${\ell}_3$ leptons ($p_{p3}=1$) or when $|{\ell}_1\rangle,
|{\ell}_2\rangle, |{\ell}_3\rangle$
form an orthonormal basis (in this case necessarily $p_{\tilde{2}_{\tilde{3}}1}=1$).

The latter would correspond to special Dirac mass matrices
corresponding a see-saw orthogonal matrix \cite{ci} that is either the identity or that
is obtained from the identity  permuting rows or columns. These special forms correspond
to so called form dominance models \cite{kingchen} and
are enforced typically by discrete flavour symmetries such as $A4$ \cite{A4}
typically justified in order to reproduce tri-bimaximal mixing \cite{tribimaximal}. However, in the limit
of exact form dominance (corresponding to the case of unbroken symmetry in flavour symmetry models)
 the total \cite{geometry} and the flavoured
$C\!P$ asymmetries vanish \cite{jenkins,leptotri}. Therefore,
a deviation from the orthonormality is necessary. For example, in models with
flavour discrete symmetries, this deviation has to be of the order of the symmetry breaking
parameter $\xi \sim 10^{-2}$ in order to generate the
correct asymmetry. However, such small deviations would still yield $w^{\rm f}\sim \xi$,
still not sufficiently small to guarantee an efficient wash-out of a large asymmetry ${\cal O}(1)$.

In the Appendix we also show the derivation of the contribution to the final asymmetry from the RH
neutrino decays, the genuine contribution from leptogenesis. However, here we just give the result
that should be regarded as a benchmark case that can be easily extended to all the other mass patterns.

The total final asymmetry is the sum of two terms
\be\label{NBmLlepTB1}
N_{B-L}^{\rm lep}(T_{B1})=N_{\Delta_1}^{\rm lep}(T_{B1})+N_{\Delta_{\tilde{1}}}^{\rm lep}(T_{B1}) \, ,
\ee
where
\bea
N_{\Delta_1}^{\rm lep}(T_{B1}) & = &
p_{21}\,p_{32}\,\ve_3\,\kappa(K_3)\,e^{-{3\pi\over 8}\,(K_1+K_2)} \\ \nonumber
& & + p_{21}\,\ve_2\,\k(K_2)\,e^{-{3\pi\over 8}\,K_1} \\ \nonumber
& & + p_{\tilde{2}_3 1}\,(1-p_{32})\,\ve_3\,\k(K_3)\,e^{-{3\pi\over 8}\,K_1} \\ \nonumber
& & + \ve_1\,\k(K_1)
\eea
and
\bea
N_{\Delta_{\tilde{1}}}^{\rm lep}(T_{B1}) & = &
(1-p_{21})\,[p_{32}\,\ve_3\,\kappa(K_3)\,e^{-{3\pi\over 8}\,K_2}+\ve_2\,\k(K_2)] \\ \nonumber
& & + (1-p_{{\tilde{2}}_3 1})\,(1-p_{32})\,\ve_3\,\k(K_3) \, .
\eea
The probabilities $p_{ij}$'s can be expressed as
\be
p_{ij}={\left|(m^{\dagger}_D\,m_D)_{ij}\right|^2
\over (m^{\dagger}_D\,m_D)_{ii}\,(m^{\dagger}_D\,m_D)_{jj}} \, .
\ee
This result is an example of how to take into account all flavour effects
in a combined way and is another main result of our paper.

%---------------------------------
\section{Light flavored scenarios}
%---------------------------------

In this section we consider RH neutrino mass patterns
where at least one $M_i$ is below $10^{12}\,{\rm GeV}$.
In this way the pre-existing lepton states, the $|{\ell}^{\rm p}\rangle$'s and the
$|{\ell}_i\rangle$'s, get partially or fully (if $M_i\ll 10^9\,{\rm GeV}$)
projected on the light neutrino flavour basis during the stages when the $N_i$ decays and inverse
processes are active.

\subsection{Scenarios with  $\tau$ flavour projection} \label{sec:light_flavor1}
%%%%%%%%%%%%%%%%%%%%%%%%%%%%%%%%%%%%%%%%%%%%%%%%%%%%%%%%%%%%%%%%%%%%%%%%%%%%%%%%

We start considering first the the three mass patterns with $M_1 \gg 10^9\,{\rm GeV}$
(see Fig. 3), where only the $\tau$ component is `measured' by the tauon Yukawa interactions.
\begin{figure}[htbp]
\begin{center}
\subfigure[]{\label{fig:step2}\includegraphics[height=5cm,width=5.2cm]{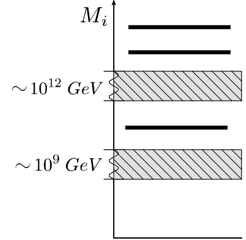}}
\hspace{10mm}
\subfigure[]{\label{fig:step3}\includegraphics[height=5cm,width=3.2cm]{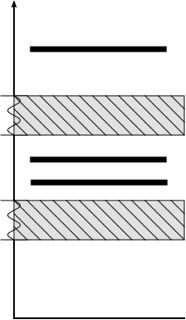}}
\hspace{10mm}
\subfigure[]{\label{fig:step4}\includegraphics[height=5cm,width=3.2cm]{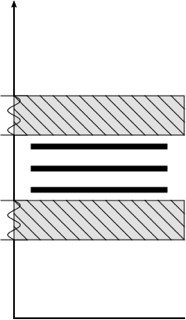}}
\end{center}
\caption{The three possible mass patterns where at least one RH neutrino mass
is comprised between $10^9\,{\rm GeV}$ and $10^{12}\,{\rm GeV}$.}
\end{figure}

\subsubsection{Case $M_{2,3}\gg 10^{12}\,{\rm GeV}$}

Let us start with the case where the two heavier RH neutrino
masses $M_{2,3}\gg 10^{12}\,{\rm GeV}$ while
$10^{12}\,{\rm GeV} \gg M_1 \gg 10^9\,{\rm GeV}$.
The evolution of the residual pre-existing asymmetry $N^{\rm p}_{B-L}$ proceeds through the same  steps
discussed in the heavy flavoured scenario until the end of the $N_2$ washout at $T \sim T_{B2}$
when it is given by the eq.~(\ref{NBmLTB2}).

At this stage an important difference arises between the two scenarios.
In the considered light flavoured scenario, before the onset of
the $N_1$ washout processes, the tauon charged lepton
interactions become effective. In this way the  $\tau$ component of the
quantum lepton states is measured and they become an incoherent mixture
of three components: a $\tau$ component, a component
${\ell}_{\tilde{\tau}_2}^{\rm p}$ (the
projection of the ${\ell}_2$ component on the $\tilde{\tau}$ plane),
and finally a component ${\ell}_{\tilde{\tau}_{\tilde{2}}}^{\rm p}$
(the projection of the ${\ell_2}$ orthogonal  component on the $\tilde{\tau}$ plane).
The residual value of the pre-existing asymmetry can be then decomposed correspondingly as
the sum of three terms,
\be\label{NpBmLTB2}
N^{\rm p}_{B-L}(10^{12}\,{\rm GeV}\gg T \gg M_1)  =
N_{{\Delta_\tau}}^{\rm p} + N_{\Delta_{\tilde{\tau}_2}}^{\rm p} + N_{\Delta_{\tilde{\tau}_{\tilde{2}}}}^{\rm p} \, ,
\ee
where
\bea
N_{{\Delta_\tau}}^{\rm p} & = &
[p_{2\tau}\,N_{\Delta_2}^{\rm p}(T_{B3})\,e^{-{3\pi\over 8}\,K_2}+
p_{\tilde{2}\tau}\,N_{\Delta_{\tilde{2}}}^{p}(T_{B3})] \\ \nonumber
N_{\Delta_{\tilde{\tau}_2}}^{\rm p} & = & (1-p_{2\tau})\,N_{\Delta_2}^{p}(T_{B3})\,e^{-{3\pi\over 8}\,K_2} \\
N_{\Delta_{\tilde{\tau}_{\tilde{2}}}}^{\rm p} & = &
 (1-p_{\tilde{2}\tau})\,N_{\Delta_{\tilde{2}}}^{\rm p}(T_{B3}) \, .
\eea
When finally the $N_1$ wash-out processes act on the pre-existing asymmetry,
one has to distinguish the wash-out acting on $N_{\Delta_\tau}^{\rm p}$, ruled by
$K_{1\tau}\equiv p_{1\tau}\,K_1$, and  the wash-out acting on $N_{\Delta_{\tilde{\tau}_2}}^{\rm p}$,
ruled by $K_{1\tilde{\tau}}\equiv (1-p_{1\tau})\,K_1$.
At the end of this stage, at $T\sim T_{B1}$,
$N_{{\Delta_\tau}}^{\rm p}$ will be therefore given by
\be
N^{\rm p}_{\Delta \tau}(T_{B1})= [p_{2\tau}\,N_{\Delta_2}^{p}(T_{B3})\,e^{-{3\pi\over 8}\,K_2}+
p_{\tilde{2}\tau}\,N_{\Delta_{\tilde{2}}}^{\rm p}(T_{B3})]\,e^{-{3\pi\over 8}\,K_{1\tau}} \, .
\ee
Imposing $K_{1\tau}\gtrsim 10$, one can enforce a strong wash-out of this component.
At the same time, imposing $K_2\gtrsim 10$, one can also enforce the wash-out of
$N_{\Delta_{\tilde{\tau}_2}}^{\rm p}$. On the other hand
the contribution $N_{\Delta_{\tilde{\tau}_{\tilde{2}}}^{\rm p}}$ does not
undergo any $N_2$ wash-out. This part of the residual pre-existing asymmetry
can be in turn decomposed as the sum of two terms:
a term $N_{\Delta_{\tilde{\tau}_{1}}}^{\rm p}$,
the asymmetry in the component of lepton states that are a projection
of $\ell_{\tilde{\tau}_{\tilde{2}}}$ on ${\ell}_{\tilde{\tau}_1}$  and
a term $N_{\Delta_{\tilde{\tau}_{\tilde{1}}}}^{\rm p}$,
the asymmetry in the states orthogonal both to ${\ell}_1$ and to ${\ell}_\tau$.
The first one is exponentially washed-out by $N_1$ inverse processes,
\be
N_{\Delta_{\tilde{\tau}_{1}}}^{\rm p}(T_{B1})=
p_{1\tilde{\tau}}\,(1-p_{\tilde{2}\tau})\,N_{\Delta_{\tilde{2}}}^{\rm p}(T_{B3})
\,e^{-{3\pi\over 8}\,(K_1-K_{1\tau})} \, ,
\ee
but the latter is not,
\be
N_{\Delta_{\tilde{\tau}_{\tilde{1}}}}^{\rm p}(T_{B1})=
(1-p_{\tilde{\tau}1})\,(1-p_{\tilde{2}\tau})\,N_{\Delta_{\tilde{2}}}^{\rm p}(T_{B3}) \, .
\ee
If we now go back to the expression eq.~(\ref{ND2TB3}) for $N_{\Delta_{\tilde{2}}}^{\rm p}(T_{B3})$,
we can see that even imposing $K_3\gtrsim 10$ there will be still a completely unwashed term
in the orthogonal $\ell_3$ component, given by
\be
N^{\rm p}_{\Delta_{\tilde{\tau}_{\tilde{1}(\tilde{2}\tilde{3})}}}(T_{B1})=
(1-p_{\tilde{\tau}1})\,(1-p_{\tilde{2}\tau})\,
(1-p_{\tilde{3}2})\,(1-p_{p3})\,N_{B-L}^{\rm p,i} \, .
\ee
Therefore, in this scenario, even imposing $K_{1\tau},K_2,K_3, (K_1-K_{1\tau}) \gtrsim 10$,
one has
\be\label{wf1}
w^{\rm f}\simeq (1-p_{\tilde{\tau}1})\,(1-p_{\tilde{2}\tau})\,
(1-p_{\tilde{3}2})\,(1-p_{p3}) \, .
\ee
This shows that in general, except for some reduction due to a geometrical projection,
there is no efficient wash-out of the pre-existing asymmetry.
Therefore, it is clear that even in this scenario
a sensible fraction of an arbitrary pre-existing asymmetry escapes
the wash-out of all three RH neutrinos. The major stages of this
scenario are summarized in Fig. 4.
\begin{figure}[htbp]
\begin{center}
\subfigure[$T\sim M_2$]{\label{fig:step2}\includegraphics[width=4.2cm]{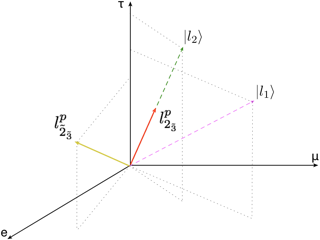}}
\hspace{5mm}
\subfigure[$T\sim 10^{12}\,{\rm GeV}$]{\label{fig:step3}\includegraphics[width=4.2cm]{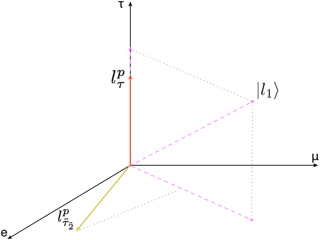}}
\hspace{5mm}
\subfigure[$T\sim M_1$]{\label{fig:step4}\includegraphics[width=4.2cm]{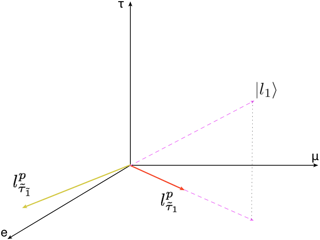}}
\end{center}
\caption{The three most significant steps in the evolution of the residual
pre-existing asymmetry in the scenario with $M_{2,3}\gg 10^{12}\,{\rm GeV}$ and
$10^{12}\,{\rm GeV} \gg M_1 \gg 10^9\,{\rm GeV}$.}
\end{figure}

\subsubsection{Case $M_{3}\gg 10^{12}\,{\rm GeV} \gg M_2 \gg M_1 \gg 10^9\,{\rm GeV}$}

It is  straightforward to extend the result of eq.~(\ref{wf1}) to a scenario
where $M_{3}\gg 10^{12}\,{\rm GeV} \gg M_2 \gg M_1 \gg 10^9\,{\rm GeV}$.
Let us then just quickly discuss a derivation of the wash-out factor
that is slightly different from the one in the previous case.

This time we have to impose $K_3\gtrsim 10$ in order to wash-out first,
at $T\sim M_3$, the component $N_{\Delta_3}^{\rm p}=p_{p3}\,N_{B-L}^{\rm p,i}$. At
$T\sim 10^{12}\,{\rm GeV}$ the lepton quantum states become an incoherent mixture
of a $\tau$ component and of a $\tilde{\tau}$ component. A condition $K_{1\tau}+K_{2\tau}\gtrsim 10$
clearly guarantees a wash-out of the asymmetry in the $\tau$ component.

At $T\sim M_2$, the lepton quantum states $\tilde{\tau}$ component
becomes a incoherent mixture of a component that is parallel
to the ${\ell}_2$ projection on the $\tilde{\tau}$ plane, that we indicate
with ${\ell}_{\tilde{\tau}_2}$, and of a component ${\ell}_{\tilde{\tau}_{\tilde{2}}}$,
the projection of the ${\ell}_2$ orthogonal component on the $\tilde{\tau}$ plane.
A condition $K_{2\tilde{\tau}}=K_{2e}+K_{2\mu}\gtrsim 10$
guarantees the  washout of the fraction of the pre-existing asymmetry in the first
component but not of the fraction in the second one.

Finally, at $T\sim M_1$, the lepton quantum states become an incoherent mixture of
a ${\ell}_{\tilde{\tau}_1}$ component and of a ${\ell}_{\tilde{\tau}_{\tilde{1}}}$ component
and again, imposing $K_{1\tilde{\tau}}=K_{1e}+K_{1\mu}\gtrsim 10$, one can enforce the wash-out
of the asymmetry in the first component but not in the second one. Therefore,
at the end there will still be a completely unwashed residual fraction of the
pre-existing asymmetry given by
\be
N_{B-L}^{\rm p,f}\simeq (1-p_{\tilde{\tau}_2\tilde{\tau}_1})\,
(1-p_{\tilde{\tau}_3\tilde{\tau}_2})\,(1-p_{\tilde{3}\tau})\,(1-p_{p3}) \,N_{B-L}^{\rm p,i} ,
\ee
showing again that the wash-out of $N_{B-L}^{\rm p,i}$ cannot be enforced in this scenario as well.

\subsubsection{Case $10^{12}\,{\rm GeV} \gg M_3 \gg M_2 \gg M_1 \gg 10^9\,{\rm GeV}$}

Finally, in the last scenario with
$10^{12}\,{\rm GeV} \gg M_3 \gg M_2 \gg M_1 \gg 10^9\,{\rm GeV}$,
the result for the final wash-out factor is a clear extension of the two
previous cases and we can directly write the final result given by
\be
w^{\rm f}\simeq (1-p_{\tilde{\tau}_{\tilde{2}}\tilde{\tau}_1})\,(1-p_{\tilde{\tau}_{\tilde{3}}\tilde{\tau}_2})\,
(1-p_{\tilde{\tau}_p\tilde{\tau}_3})\,(1-p_{p\tau}) \, ,
\ee
showing that in general, even in this case, $N_{B-L}^{\rm p,i}$ cannot be completely washed-out.

Therefore, we can conclude that in all mass patterns with $M_1\gg 10^{9}\,{\rm GeV}$
it is not possible to enforce an efficient washout of a large pre-existing asymmetry.

\subsection{Scenarios with  $M_1\ll 10^{9}\,{\rm GeV}$}
%%%%%%%%%%%%%%%%%%%%%%%%%%%%%%%%%%%%%%%%%%%%%%%%%%%%%%%

Let us now discuss those
mass patterns where at least one $M_i\ll 10^9\,{\rm GeV}$.
There are 6 different possibilities, as sketched in Fig. 5.
\begin{figure}[htbp]
\begin{center}
\subfigure[]{\label{fig:step2}\includegraphics[height=3cm,width=4cm]{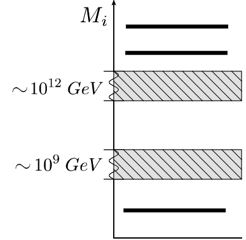}}
\hspace{10mm}
\subfigure[]{\label{fig:step3}\includegraphics[height=3cm,width=2.5cm]{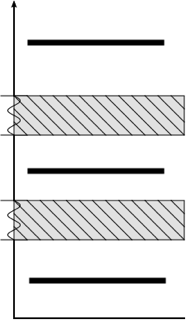}}
\hspace{10mm}
\subfigure[]{\label{fig:step4}\includegraphics[height=3cm,width=2.5cm]{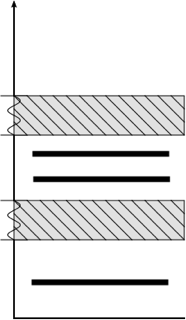}} \\
\subfigure[]{\label{fig:step2}\includegraphics[height=3cm,width=4cm]{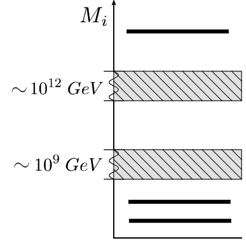}}
\hspace{10mm}
\subfigure[]{\label{fig:step3}\includegraphics[height=3cm,width=2.5cm]{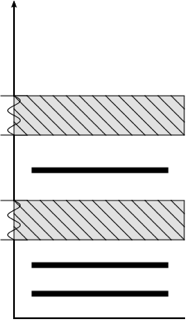}}
\hspace{10mm}
\subfigure[]{\label{fig:step4}\includegraphics[height=3cm,width=2.5cm]{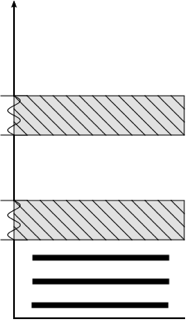}}
\end{center}
\caption{The six possible mass patterns with $M_1\ll 10^9\,{\rm Gev}$.
Only (b) and (c) allow for a successful strong thermal leptogenesis. }
\end{figure}
It is quite clear, from our previous discussions, that for all them it is always possible
to enforce a strong wash-out of the pre-existing asymmetry imposing
$K_{1e}, K_{1\m}, K_{1\t} \gtrsim 10$ \cite{nir}.
Indeed if at $T\sim T_{B2}$, when the $N_2$-inverse processes wash-out freeze, the residual
value of the pre-existing asymmetry is given by $N_{B-L}^{\rm p}(T\sim T_{B2})$, then
at $T \sim M_1 \ll 10^9\,{\rm GeV}$, irrespectively of the value of $T_{B2}$, this will
be distributed in leptons and anti-leptons quantum states that are an incoherent mixture
of the three light neutrino flavours. Therefore, the $N_1$ wash-out will act separately on each
flavour contribution $N^{\rm p}_{\Delta_{e,\m,\t}}$ to the total residual pre-existing
asymmetry $N_{B-L}^{\rm p}(T_{B2})$. In this way,
at the end of the $N_1$ washout, the final value of the
residual fraction of the pre-existing asymmetry is given by
\be\label{NBmLN2DS}
N_{B-L}^{\rm p,f}= \sum_{\a=e,\m,\t}\, N^{\rm p}_{\Delta_{\a}}\,e^{-{3\pi\over 8}\,K_\alpha} \, .
\ee
Therefore, imposing $K_{1e}, K_{1\m}, K_{1\t} \gtrsim 10$, one can this time always enforce
a sufficiently strong washout of a pre-existing large asymmetry ($w^{\rm f}\lesssim 10^{-8}$).
However, such a strong condition would also wash-out the contribution $N_{B-L}^{\rm lep,f}$
produced from the two heaviest RH neutrino decays while on the other hand,
the lightest RH neutrino $C\!P$ asymmetries are
too low to have enough asymmetry produced from the $N_1$ decays \cite{di}.
Therefore, this condition  is incompatible with successful leptogenesis
\footnote{There is a loophole. In \cite{bounds} it was shown
that the flavoured $C\!P$ asymmetries contain a term that is not upper bounded
if one allows for strong cancellations in the light neutrino masses from the seesaw formula
corresponding to large entries of the orthogonal matrix. Successful leptogenesis
from $N_1$ decays is then possible for $M_1\ll 10^9\,{\rm GeV}$.}.

We have then to find a weaker condition that can satisfy simultaneously
$w^{\rm f}\lesssim 10^{-8}$ and $N_{B-L}^{\rm lep,f} \sim 10^{-7}$ or,
using a vivid analogy, the way not to throw the baby out with the bath water.
It is clear that at least one of the $K_{1\a}$ has to be $\lesssim 1$.
The wash-out of the pre-existing asymmetry has then to be carried out by the
heavier RH neutrinos and their decays have also  to produce  $N_{B-L}^{\rm lep,f}$
in the $\a$ flavour after the freeze-out of the wash-out processes.

\subsubsection{Case $M_{2,3} \gg 10^{12}\,{\rm GeV}$}

Let us start from a mass pattern where $M_{2,3}\gg 10^{12}\,{\rm GeV}$.
At $T\sim T_{B2}$ the residual value of the pre-existing asymmetry is
given by the eq.~(\ref{NBmLTB2}) and, imposing $K_2\gtrsim 10$, only
the asymmetry in the ${\ell}_{\tilde{2}}$ states survives
(the second term in the eq.~(\ref{NBmLTB2})). At $T\sim 10^{12}\,{\rm GeV}$,
all lepton quantum states become an incoherent mixture of a $\tau$ component
and of a $\tilde{\tau}$ component. The asymmetry produced from the $N_2$ decays
at $T\sim T_{B2}$ is, by definition, all in the ${\ell}_2$ flavour, i.e.
$N_{B-L}^{\rm lep}(T_{B2})=N_{\D_{\ell_2}}^{\rm lep}(T_{B2})$.

Below $T\sim 10^{9}\,{\rm GeV}$, both the two contributions
to the asymmetry in  states in the $\tilde{\tau}$ plane,
the residual pre-existing fraction and the one produced by RH neutrino decays,
will be distributed in quantum lepton states that are
an incoherent mixture of a muon and of an electron component. Therefore, there is a
residual fraction of the pre-existing asymmetry in each light lepton flavour.
This implies that it is impossible to impose a condition such that all
the residual pre-existing asymmetry is washed-out without also washing out
the contribution produced from RH neutrino decays.

\subsubsection{Case $10^{12}\,{\rm GeV}\gg M_2 \gg 10^{9}\,{\rm GeV}$}

Finally, we consider a scenario with $M_1\ll 10^{9}\,{\rm GeV}$ and
$10^{12}\,{\rm GeV}\gg M_2 \gg 10^{9}\,{\rm GeV}$.
At $T\sim M_2$, the lepton quantum states become an incoherent mixture of
a $\tau$ component and of a $\tilde{\tau}$ component.
We can again impose $K_{2\tau}\gtrsim 10$ in such a way that any residual
pre-existing asymmetry $\lesssim {\cal O}(1)$ in the tauon flavour is washed-out.
However, this time the out-of-equilibrium
$N_2$-decays can still afterwards produce an asymmetry in the same tauon
flavour sufficient to have successful leptogenesis. The asymmetry has necessarily to be
produced in the tauon flavour since we have still to enforce a strong wash-out
of the pre-existing asymmetry in the electron and muon flavours from the
$N_1$ inverse processes by imposing $K_{1e},K_{1\m} \gg 1$. Notice that
successful leptogenesis requires $\ve_{2\tau}\sim 10^{-6}$, so that at the end $\eta_B \sim
\ve_{2\tau}\,\kappa(K_{2\tau}) \sim 10^{-9}$. In this way
we have finally pinned down a scenario that can realize successful
strong thermal  leptogenesis: {\em a tauon $N_2$-dominated scenario}
\footnote{Notice that in a tauon $N_2$-dominated scenario with $K_{2\tau}\gg 1$ (strong wash-out regime)
one has an expression for the final asymmetry that is basically given by the expression
valid in the traditional unflavoured $N_1$ dominated scenario where $K_1$ is replaced by $K_{2\tau}$.
In this case one obtains (see for example right panel of Fig.~1 in \cite{flavorlep})
 an upper bound $K_{2\tau}\ll 1000$ from the condition $M_2\ll 10^{12}\,{\rm GeV}$}.
It is interesting to notice that in this scenario the presence of a third RH neutrino specie $N_3$ is necessary for $\ve_{2\tau}$ not to be suppressed as
$\sim (M_1/M_2)\,10^{-6}\,{M_1/10^{10}\,{\rm GeV}}$.

\begin{figure}[htbp]
\begin{center}
\subfigure[$T \sim T_{B3}$]{\label{fig:step1}\includegraphics[width=7.2cm]{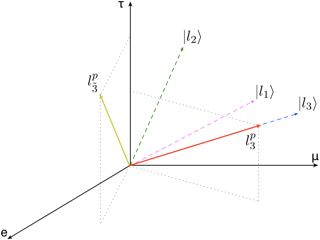}}
\subfigure[$T\sim 10^{12}\,{\rm GeV}$]{\label{fig:step2}\includegraphics[width=7.2cm]{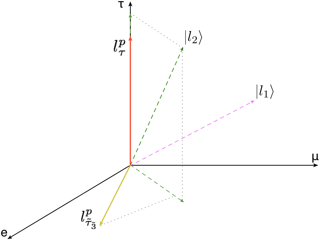}}\\
\subfigure[$T\sim T_{B2}$]{\label{fig:step3}\includegraphics[width=7.2cm]{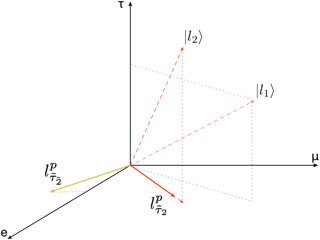}}
\subfigure[$T\sim T_{B1}$]{\label{fig:step4}\includegraphics[width=7.2cm]{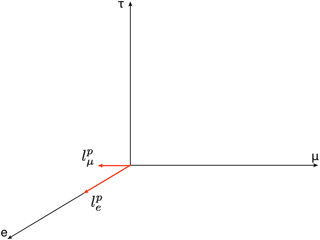}}\\
\end{center}
\caption{Tauon
$N_2$ dominated scenario.
Steps in the evolution of the lepton quantum states carrying the residual pre-existing asymmetry.
As in the previous figures, the red components are those undergoing the wash-out,
while the yellow component escapes the wash-out.
The yellow component, tracking the lepton  states carrying
an unwashed fraction of the pre-existing asymmetry, is not present
after the $N_1$ wash-out at $T\sim M_1$. We have finally singled out the only
mass pattern where successful strong thermal leptogenesis is possible.}
\end{figure}
It is easy to understand that there cannot be a scenario of successful strong thermal
leptogenesis where the final asymmetry is dominantly in the electron or in the muon flavour.
Suppose that we impose $K_{2e}+K_{2\mu}\gtrsim 10$, so that all the pre-existing asymmetry
in the $\tilde{\tau}_2$ component is washed out at $T\sim M_2$.
Suppose that afterwards a sufficiently high asymmetry is generated in the
$\tilde{\tau}_2$ component by out-of-equilibrium $N_2$ decays at $T\sim T_{B2}$.
However, there is still a $\tilde{\tau}_{\tilde{2}}$ component
$N_{\Delta_{\tilde{\tau}_{\tilde{2}}}}^p(T_{B2})$, that escapes the wash-out.

Indeed at $T\ll 10^{9}\,{\rm GeV}$, the lepton quantum states
become an incoherent mixture of an electron and of a muon component. If we
impose $K_{1\tau}+K_{2\tau} \gtrsim 10$ we can wash-out efficiently the residual
pre-existing asymmetry in the tauon flavour. However, either $K_{1e}$
or $K_{1\m}$, has to be necessarily  $\lesssim 1$ otherwise everything,
including $N^{\rm lep}_{B-L}$, would be
washed-out. For definiteness let us choose $K_{1e}\lesssim 1$. This implies
that there will be still a final residual value of the asymmetry in the electron
flavour given by
\be
N_{\D_e}^{f,p}=p_{\tilde{\tau}_{\tilde{2}}e}\,N_{\Delta_{\tilde{\tau}_{\tilde{2}}}}^p(T_{B2})
\ee
that cannot be washed-out. Of course the same  would happen if we would impose
$K_{1\m}\lesssim 1$ instead of $K_{1e}\lesssim 1$.

Notice that if we consider a mass pattern where both $M_2$ and $M_3$
are in the range $[10^9,10^{12}]\,{\rm GeV}$, then things would work exactly in the same
way. Simply in this case one can impose the less restrictive condition $K_{2\tau}+K_{3\tau}\gtrsim 10$
in order to wash-out the pre-existing asymmetry in the tauon flavour.
Notice that a tauon $N_3$-dominated scenario could be  in principle possible
if $K_{3\tau}\gtrsim 10$ and $K_{2\tau}\lesssim 1$.
However, the maximal value of $\ve_{3\tau}$ is suppressed as $\propto M_2/M_3$ with
respect to $\ve_{2\tau}$ and therefore the asymmetry produced from $N_2$ decays
tends to be much higher both because of the lower wash-out and because of the
much higher $C\!P$ asymmetry. Therefore, this possibility can be
possibly realized only for a very fine tuned choice of the parameters and
in any case not for a too strong hierarchy between $M_2$ and $M_3$.
For this  reason mass patterns with $M_2\ll 10^{9}\,{\rm GeV}$ and $M_3\gg 10^9\,{\rm GeV}$
cannot give successful leptogenesis.

In this way we have  finally shown that the only possible scenario
with a hierarchical RH neutrino mass spectrum which allows for
successful strong thermal leptogenesis is a tauon $N_2$-dominated scenario,
where $M_1\ll 10^{9}\,{\rm GeV}$ and $10^{12}\,{\rm GeV}\gg M_2 \gg 10^9\,{\rm GeV}$
and the final asymmetry is dominantly in the tauon flavour. In Fig. 6 we have sketched the
4 main steps in the evolution of the flavour composition of $N_{B-L}^{\rm p}$.

%%%%%%%%%%%%%%%%%%%%%%%%%%%%%%%%%%%%%%%%%%
\section{Remarks, Caveats and Conclusions}
%%%%%%%%%%%%%%%%%%%%%%%%%%%%%%%%%%%%%%%%%%

Let us make some remarks and indicate some caveats about the results that we have found,
in particular pointing out the limits of validity and some alternative ways to realize
successful strong thermal leptogenesis relaxing some of the assumptions made so far.

\subsection{Beyond the hierarchical limit}

Our conclusions have been derived within the hierarchical limit. If even just the two lightest RH neutrinos
masses are allowed to be almost-degenerate and lower than $10^{9}\,{\rm GeV}$,
then the $C\!P$ asymmetries are enhanced \cite{crv} and successful leptogenesis is possible.
At the same time, the asymmetry from the RH neutrino decays is produced
in the three flavour regime. Therefore, before the onset of leptogenesis, a
pre-existing asymmetry gets distributed
into quantum lepton states that are an incoherent mixture of all
three light flavour eigenstates. In this case, as already discussed, an efficient wash-out
can be enforced imposing $K_{1\a}\gtrsim 10$ for all $\a$ without spoiling successful leptogenesis.
Therefore, this is a clear loophole in our conclusions. This agrees with previous results
obtained within the context of resonant tau-leptogenesis \cite{resonant}.

Less trivially, it should be noticed that a scenario where all three RH neutrino masses are quasi-degenerate
should realize successful strong thermal leptogenesis for any value of $M_i$
(e.g. even in the case of  the heavy flavoured scenario of Fig. 1).
 Our results were obtained under the assumption that the wash-out from different RH neutrinos
inverse processes occur at different stages. In this way the wash-out of the
asymmetry occurs first along one direction, e.g. $\ell_3$, and then along a
different one, e.g. ${\ell}_2$. This implies that at any stage there is  a well
defined flavour basis where the density matrix is diagonal. If the wash-out occurs simultaneously along
three different independent directions, even not orthogonal to each other, then
a  pre-existing asymmetry should be efficiently washed-out anyway. However, this
heuristic argument would require to be more rigorously
proved within a density matrix formalism.

\subsection{Supersymmetric models}

In the case of supersymmetric models, charged lepton interactions can become
considerably faster and the condition eq.~(\ref{cond3})
for the full projection of the leptonic quantum states on
the light flavor basis gets relaxed to \cite{susy}
\be
M_i \ll 10^{9}\,{\rm GeV}\,(1+\tan^2\beta) .
\ee
On the other hand, the lower bound on the RH neutrino mass
for successful leptogenesis in the strong wash-out regime,
$M_1\gtrsim 2\times 10^{10}\,{\rm GeV}\,(K_{1\a}/10)^{1.2}$,
does not change significantly \cite{lbsusy,flavorlep}.
Therefore, for $1+\tan^2\beta\gtrsim 20$ , it is possible to have a full wash-out of the
pre-existing asymmetry imposing $K_{1e}, K_{1\mu}, K_{1\t}\gtrsim 10 $ and at the
same time successful leptogenesis from light RH neutrino decays. In this way
a traditional $N_1$-dominated leptogenesis scenario where the asymmetry
is produced in the three flavour regime is possible.
Of course it is also still possible to realize successful strong thermal leptogenesis within a $N_2$-dominated scenario,
for $K_{2\a}\gtrsim 10$ and $K_{1\a}\lesssim 1$, this time in an arbitrary flavour $\a$.

\subsection{Light neutrino flavour coupling}

Light flavour asymmetries do not evolve independently of each other, and in
a $N_2$-dominated scenario, light flavour coupling can play a relevant role \cite{flcoup}.
In our case the main issue is whether flavour coupling can spoil
the efficient wash-out of the pre-existing asymmetry in the tauon $N_2$-dominated scenario
allowing the electron and muon flavour asymmetries
to escape the wash-out in general. This is because the eq.~(\ref{NBmLN2DS})
for the final total asymmetry
has now to be replaced by a much more general equation for the final individual
flavour contributions to the asymmetry
\bea \nonumber
N^{\rm f}_{\D_{\a}} & = & \sum_{\a''}\,V^{-1}_{\a\a''}\,
\left[N^{T\sim T_{B2}}_{\a''}\,e^{-{3\pi\over 8}\,K_{1\a''}}\right] \, ,
\eea
where $V_{\a\a''}$ is a mixing matrix that can be derived from the flavour
coupling mixing matrix and the $N^{T\sim T_{B2}}_{\a''}$'s are a linear
combination of the $N^{T\sim T_{B2}}_{\a}$'s. In first approximation the
matrix $V$ is diagonal and one recovers the eq.~(\ref{NBmLN2DS}) but
in general one can see how the wash-out for example of the the $\m$ contribution
can partly proceed through an exponential containing $K_{1\t}$. Therefore, in principle, this
component can escape the wash-out in the tauon $N_2$-dominated scenario.
However, as one can see, the result depends in a complicated way on the seesaw parameters
and it should be checked at each point in the seesaw parameter space
whether a pre-existing asymmetry is efficiently washed out or not.
Light neutrino flavour coupling could therefore potentially introduce further
restrictions on the tauon $N_2$-dominated scenario.

\subsection{Heavy neutrino flavour coupling}

As for the light neutrino flavour asymmetries, the dynamics of the  heavy neutrino flavour asymmetries
also couple when spectator processes are considered \cite{nir}.
For example, in the heavy flavoured scenario, for $K_3\gtrsim 10$, one would have that the
$N_{\Delta_3}=-N_{\Delta_{\tilde{3}}}/4$, implying that part of the component
of pre-existing asymmetry in the ${\ell}_3$ quantum states is not completely washed out.
This result can be understood considering that when  the Higgs asymmetry $N_H$
is taken into account  in the kinetic equation for $N_{\Delta_3}$ (cf. eq.~(\ref{heavyflke})),
\be
{dN_{\D_3}^{\rm p}\over dz_3}  =  -W_3\,\left(N_{\Delta_3}^{\rm p}+{N_{H}\over 2}\right) \, ,
\ee
it couples the wash-out of $N^{\rm p}_{\Delta_3}$ to the value of $N^{\rm p}_{\Delta_{\tilde{3}}}$.
At the same time the Higgs asymmetry is enforced, by equilibrium of gauge and top Yukawa
interactions plus baryon number and hypercharge conservation,
to be $N_H=N_{\Delta_{\tilde{3}}}/2$. However, an account of this effect would just introduce
${\cal O}(1)$ corrections to the wash-out factors without changing our conclusions.

\subsection{An improved description of decoherence}

As we have seen decoherence of lepton quantum states originates from two different interactions:
the $N_i$ inverse processes that tend to collapse the quantum states into a incoherent
mixture of a ${\ell}_i$ component and of a ${\ell}_{\tilde{i}}$ component and from
charged lepton interactions that, for example, at $T\sim 10^{12}\,{\rm GeV}$
tend to collapse the lepton quantum states into a $\t$ component and into a $\tilde{\tau}$
component. In the first case a description of decoherence in terms of density matrix equations
cannot change our conclusions. It can only just slightly affect
the precise value of $K_{\star}$, i.e. the minimum value for the $K_i$'s and the $K_{i\a}$'s to get an efficient
wash-out. On the other hand, in the second case it would be important
to have an exact description in order to get the precise conditions on the $M_i$'s
in order for the different mass patterns to hold,
in particular those ones for the $N_2$ dominated scenario. These conditions
have been worked out in the calculation of $N^{\rm lep,f}_{B-L}$ \cite{dr}
but they could be slightly different in the calculation of $N^{\rm p,f}_{B-L}$.

\subsection{Phantom terms and dependence on the initial RH neutrino abundance}

We have so far assumed that the flavour compositions of the pre-existing leptons
and anti-leptons (cf. eqs.~(\ref{flavcomprlep}) and (\ref{flavcomprantilep}))
are the same. If we relax this assumption, we have to take into account the so called
phantom terms \cite{flcoup} that give an additional
way to a pre-existing asymmetry to avoid the wash-out in all scenarios
considered but not in the tauon $N_2$-dominated scenario.
The phantom terms
give a contribution to the flavoured asymmetries coming from possible differences in the
flavor composition of the pre-existing lepton
and anti-lepton quantum states that do not vanish  when the total
number of leptons and of anti-leptons are the same.

The flavour composition of leptons and anti-leptons are respectively given by
the eq.~(\ref{flavcomprlep}) and by the eq.~(\ref{flavcomprantilep}). We have so far assumed
that ${\cal C}_{p\a}=\bar{\cal{C}}_{p\a}$. Let us now drop this assumption.
Let us first give as an example what happens at the end of the first stage,
at temperatures $T \sim M_3 $. The pre-existing flavour asymmetries can be again decomposed as
in the eq.~(\ref{decomposition}) but  this time the terms
$N_{\D_3}^{\rm p,i}$ and $N_{\D_{\tilde{3}}}^{\rm p,i}$ are the sum of two terms,
\be
N_{\Delta_3}^{\rm p,i}= {p_{p3}+\bar{p}_{p3}\over 2}\,N_{B-L}^{\rm p,i}
                       +{p_{p3}-\bar{p}_{p3}\over 2}\,(N_{{\ell}^{\rm p}}+N_{\bar{\ell}^{\rm p}}) .
\ee
and
\be
N_{\D_{\tilde{3}}}^{\rm p,i}= {p_{p\tilde{3}}+\bar{p}_{p\tilde{3}}\over 2}\,N_{B-L}^{\rm p,i}
                       +{\bar{p}_{p\tilde{3}}-\bar{p}_{p\tilde{3}}\over 2}\,
                       (N_{{\ell}^{\rm p}}+N_{\bar{\ell}^{\rm p}})\, .
\ee
The first terms are the usual contributions coming from a difference in the number of
leptons and antileptons, the $N_{B-L}^{\rm p,i}$ asymmetry, that gets shared between the two
flavours just proportionally to the average of the two probabilities that the lepton is
in one of the two flavour.
The second terms are the so called phantom terms \cite{flcoup} and are due
to a different flavour composition of the lepton and anti-lepton quantum states.
They are proportional to the number of leptons (in thermal equilibrium one has simply
$N_{{\ell}^{\rm p}}\simeq N_{\bar{\ell}^{\rm p}} \simeq 1$).
Notice that the sum of the two phantom terms vanishes at $T\sim M_3$ since
$p_{p3}+p_{p\tilde{3}}=\bar{p}_{p3}+\bar{p}_{p\tilde{3}}=1$.
The phantom terms are such that when the associated flavour components (in the example ${\ell}_3$ and ${\ell}_{\tilde{3}}$)
are in a coherent superposition, then they cannot be washed-out. This is because the wash-out is just
a statistical damping of the difference between the number of lepton and of anti-leptons
of the incoherent components. This means that these terms escape the wash-out exactly
like the orthogonal components but with the difference that their sum vanish. However,
in a later stage, when the phantom terms flavour components are also measured, they
can be finally washed out. If the washout proceeds in a asymmetric way, their sum does
not vanish any more and the phantom terms would finally show up.

Let us give a second example that well illustrates this last statement.
Consider the mass pattern in the left panel
of Fig. 3. At the decoherence at $T\sim 10^{12}\,{\rm GeV}$,
if phantom terms are neglected, the $N_{\Delta_{\tau}}$ asymmetry is given by the
first term in the RH side of eq.~(\ref{NpBmLTB2}). If phantom terms are taken into account
there would be an additional contribution given by
\be
N_{\Delta_{\tau}}^{\rm p,phantom}={p_{p\tau}-\bar{p}_{p\tilde{\tau}}\over 2}\,
(N_{{\ell}_2^{\rm p}}+N_{\bar{\ell}_2^{\rm p}}) \, ,
\ee
and an opposite term $N_{\Delta_{\tilde{\tau}}}^{\rm p, phantom}$.
Notice again that phantom terms are proportional to the total number of pre-existing leptons.
At the lightest RH neutrino wash-out stage, at $T\sim M_1$, if for example $K_{1\tau}\lesssim 1$ and $K_{1\tilde{\tau}}\gtrsim 10$,
then all the asymmetry $N_{\Delta_{\tilde{\tau}}}^{\rm p, phantom}$ is efficiently washed-out but the asymmetry
$N_{\Delta_{\tau}}^{\rm p,phantom}$ survives.

Analogously one would have phantom terms in the electron and in the muon flavour in the
$N_2$ dominated scenario after the decoherence stage at $T\sim 10^{9}\,{\rm GeV}$.

It is however clear that an account of the phantom terms
cannot spoil our conclusion about the tauon $N_2$ dominated scenario, i.e. they
are also efficiently washed out. This is because
when the electron and muon components become an incoherent mixture at $T\sim 10^{9}\,{\rm GeV}$,
even though now one can have phantom terms that survived the previous
$N_2$ wash-out along ${\ell}_{\tilde{\tau}_2}$, these are anyway efficiently washed out, together with the component
along ${\ell}_{\tilde{\tau}_{\tilde{2}}}$, by the $N_1$ wash-out if $K_{1e}, K_{1\m}\gtrsim 10$.

There is another aspect concerning the role of phantom terms. There are also phantom terms
 directly in the asymmetries
generated by the $RH$ neutrinos, due to a different flavour composition of the
quantum states $|{\ell}_i\rangle$ with respect to the flavour composition of the
$C\!P$ conjugated $|\bar{\ell}'_i\rangle$ \cite{flcoup}. These phantom terms
introduce an additional source of dependence on the initial abundance of RH neutrinos.
However, again, their presence can only worsen the dependence on the initial conditions for all
scenarios but the tauon $N_2$ dominated scenario. This is because
also these phantom terms are efficiently washed out by the $N_1$
inverse processes if $K_{1e}, K_{1\m}\gtrsim 10$.
Therefore, if $K_{2\tau}\gtrsim K_{\star}$  \cite{flavorlep}, we can conclude that the tauon
$N_2$-dominated scenario is fully independent of the initial conditions,
both of an initial pre-existing asymmetry and of the initial RH neutrino abundance.

\subsection{On the leptogenesis conspiracy}

Notice that the usual observation that
strong thermal leptogenesis is possible because the solar and the atmospheric neutrino
mass scales are found to be just about one order of magnitude bigger than the equilibrium neutrino mass
$m_{\star}$, typically implying $K_i=10-50$, still holds now that flavour effects are taken into account.
It is indeed still a necessary condition since otherwise it would not be possible to impose
$K_{2\tau}+K_{3\tau},K_{1e},K_{1\mu} \gtrsim 10$ without fine tuning. Simply, as we have seen, when
 flavour effects are taken into account, additional conditions have to be
 imposed.

\subsection{On the relevance of the requirement of strong thermal leptogenesis}

One could legitimately just ignore the problem of the initial conditions,
for example assuming conservatively vanishing RH neutrino abundances and vanishing pre-existing asymmetry.
However, there are many reasons to think that at the end of inflation, or in any case before leptogenesis,
the Universe was already in a $C\!P$ non invariant state \cite{affleckdine,gravity,GUTB}. It would be therefore
quite important, in this perspective, to investigate further these possible  mechanisms of production of a
a pre-existing asymmetry before the onset of leptogenesis. If it could be possible to conclude
that a large ${\cal O}(1)$ asymmetry unavoidably emerges after inflation, then
strong thermal leptogenesis would become a necessary unescapable requirement.
However, even though at the moment we lack such a stringent motivation, any model that yields solutions
respecting the strong thermal leptogenesis requirement should be regarded certainly more attractive,
and, on more practical grounds, any kind of prediction that can be possible derived from it as more robust.

\subsection{Conclusions}

We have seen how a full account of heavy and light neutrino flavour effects
gives many ways to a pre-existing asymmetry to escape the wash-out of the RH neutrinos.
On the other hand, it is quite interesting that there is
a well defined scenario, the tauon $N_2$ dominated scenario, where
successful strong thermal leptogenesis is possible. If, to a first sight,  flavour effects can
therefore seem to spoil the attractiveness of thermal leptogenesis, it is also true
that thanks to flavour effects the observed asymmetry
can be successfully produced by the next-to-lightest RH neutrinos in quite a natural way,
so that the lower bound on $M_1$ is basically nullified. In this
way, models emerging  from grand-unified theories, such as so called `$SO(10)$-inspired
models' \cite{lepgut,SO10}, can well explain the observed asymmetry. It is then quite
intriguing that in these models the final asymmetry is dominantly produced just in the tauon flavour and that all
conditions for successful strong thermal leptogenesis that we have derived in this paper are fulfilled.
It is also greatly interesting that these models seem to give rise to predictions on the
low energy neutrino parameters that can be tested in future experiments during next years,
opening new hopes to a testability of the minimal seesaw mechanism in combination with
leptogenesis. In conclusion, from our analysis,  the  tauon $N_2$ dominated scenario
emerges  as a particularly attractive realization of leptogenesis.

\vspace{3mm}

\textbf{Acknowledgments}

PDB and LM acknowledge financial support from the NExT Institute and from SEPnet.
PDB wishes to thank Ferruccio Feruglio and Steve King for useful discussions.

%%%%%%%%%%%%%%%%%%%
\section*{Appendix}
\appendix
\renewcommand{\thesection}{\Alph{section}}
\renewcommand{\thesubsection}{\Alph{section}.\arabic{subsection}}
\def\theequation{\Alph{section}.\arabic{equation}}
\renewcommand{\thetable}{\arabic{table}}
\renewcommand{\thefigure}{\arabic{figure}}
\setcounter{section}{1}
\setcounter{equation}{0}
%%%%%%%%%%%%%%%%%%%

\subsection*{Heavy Flavoured scenario}

\subsubsection*{The heavy flavour basis}

Given the Dirac mass matrix, the probabilities $p_{ij}$'s
can be expressed as
\be
p_{ij}={\left|(m^{\dagger}_D\,m_D)_{ij}\right|^2
\over (m^{\dagger}_D\,m_D)_{ii}\,(m^{\dagger}_D\,m_D)_{jj}} \, .
\ee
In a basis where the charged lepton and the Majorana mass matrices are
diagonal, the Dirac mass matrix can be parameterized in terms of the
leptonic mixing matrix $U$ and of the orthogonal matrix $\Omega$ \cite{ci}
as $m_D=U\,\sqrt{D_m}\,\Omega\,\sqrt{D_M}$.
From the unitarity of $U$, it easily follows that
\be
p_{ij}={|\sum_h\,m_h\,\O^{\star}_{hi}\,\O_{hj}|^2 \over \mti\,\mtj } \, .
\ee
In general $p_{ij}\neq \d_{ij}$
(i.e. in general the heavy flavoured basis is not orthonormal) but for
$\O$ equal to the unity matrix or to the five special forms obtained from the unity
matrix permuting rows and columns it can be indeed verified,
that $p_{ij} = \d_{ij}$.
These special forms correspond to so called form dominance models \cite{kingchen}.
However, for these 6 special cases the total $C\!P$ asymmetries vanish and successful
leptogenesis is not attained \cite{geometry} since
RH neutrinos do not interfere ($(m^{\dagger}_D\,m_D)_{i\neq j}=0$).

\subsubsection*{Wash-out factor}

A full expression for the residual value of the
pre-existing asymmetry in the heavy flavoured scenario
can be written as the sum of three terms
 \be\label{NBmLpfhf2}
N_{B-L}^{\rm p,f} = N_{\Delta_1}^{\rm p,f}(T_{B1})+
                    N_{\Delta_{\tilde{1}_2}}^{\rm p,f}(T_{B1})+
                    N_{\Delta_{\tilde{1}_{\tilde{2}}}}^{\rm p,f}(T_{B1}) \, .
\ee
The first term is the contribution to the residual value of the pre-existing asymmetry
in ${\ell_1}$ lepton quantum states and is given by
\bea
N_{\Delta_1}^{\rm p,f}(T_{B1}) & = &
N_{B-L}^{\rm p,i}\,\left[p_{21}\,p_{32}\,p_{p3}\,e^{-{3\pi\over 8}(K_1+K_2+K_3)} \right. \\ \nonumber
                               &  & + p_{21}\,p_{\tilde{3}2}\,(1-p_{p3})\,e^{-{3\pi\over 8}\,(K_1+K_2)} \\ \nonumber
                               &  & + p_{\tilde{2}_3 1}\,(1-p_{32})\,p_{p3}\,e^{-{3\pi\over 8}\,(K_1+K_3)} \\ \nonumber
                               &  & \left. + p_{\tilde{2}_{\tilde{3}}1}\,(1-p_{\tilde{3}2})\,(1-p_{p3})\,e^{-{3\pi\over 8}\,K_1}
                               \right]\, .
\eea
The second term is the contribution in ${\ell}_{\tilde{1}_2}$ lepton quantum states given by
\bea
N_{\Delta_{\tilde{1}_2}}^{\rm p,f}(T_{B1}) & = &  N_{B-L}^{\rm p,i}\,\left[
(1-p_{21})\,p_{32}\,p_{p3}\,e^{-{3\pi\over 8}\,(K_2+K_3)} \right .\\ \nonumber
                               &  &
                              \left.  + (1-p_{21})\,p_{\tilde{3}2}\,(1-p_{p3})\,e^{-{3\pi\over 8}\,K_2}\right] \, .
\eea
Finally, the third term is the contribution in ${\ell}_{{\tilde{1}_{\tilde{2}}}}$ lepton quantum states given by
\bea
N_{\Delta_{\tilde{1}_{\tilde{2}}}}^{\rm p,f}(T_{B1}) & = &
                            N_{B-L}^{\rm p,i}\,
                            \left[(1-p_{\tilde{2}1})\,(1-p_{32})\,p_{p3}\,e^{-{3\pi\over 8}\,K_3} \right. \\ \nonumber
                               &  &\left. + (1-p_{\tilde{2}_{\tilde{3}}1})\,(1-p_{\tilde{3}2})\,(1-p_{p3})\right].
\eea
Imposing $K_1, K_2, K_3 \gg 1$ one can wash-out all terms except the last
one, obtaining for the wash-out factor the expression (\ref{wfhf}).
It can be checked that the sum of all 8 probabilities adds to unity.

\subsubsection*{Asymmetry from leptogenesis}

Let us give the expression for $N_{B-L}^{\rm lep}$ in the three different stages
of the heavy flavoured scenario (it is the most involved case, extension to other
scenarios is straightforward) assuming $T_i\gg M_3$ and neglecting heavy flavour
coupling \cite{nir}.

At the end of the $N_3$ wash-out stage the asymmetry produced from $N_3$
decays in the ${\ell}_3$ states (from the eqs. (\ref{heavyflke})) is simply given by
\be
N_{B-L}^{\rm lep}(T_{B3})=N_{\Delta_3}^{\rm lep}(T_{B3})=\ve_3\,\kappa(K_3) \, .
\ee
At the end of the $N_2$ wash-out stage, the asymmetry can be written
as the sum of two components,
\be\label{NBmLlepTB2}
N_{B-L}^{\rm lep}(T_{B2})=N_{\Delta_2}^{\rm lep}(T_{B2})+N_{\Delta_{\tilde{2}_3}}^{\rm lep}(T_{B2}) \, .
\ee
The first components is the asymmetry in ${\ell}_2$  states.
This is washed out by $N_2$ inverse processes
and  is in turn the sum of two terms,
\be
N_{\Delta_2}^{\rm lep}(T_{B2})=
p_{32}\,\ve_3\,\kappa(K_3)\,e^{-{3\pi\over 8}\,K_2}+\ve_2\,\k(K_2) \, ,
\ee
the first one is the asymmetry produced by $N_3$ decays while the second one is the
asymmetry produced by $N_2$ decays.

The second term in the eq.~(\ref{NBmLlepTB2}), in ${\ell}_{\tilde{2}_{3}}$ quantum states,
is the asymmetry  produced by $N_3$ decays that is not washed-out by $N_2$ inverse
processes and can be written as
\be
N_{\Delta_{\tilde{2}_3}}^{\rm lep}(T_{B2})=(1-p_{32})\,\ve_3\,\k(K_3) \, .
\ee
At the end of the $N_1$ wash-out we can again write the total asymmetry as the sum of two terms
\be\label{NBmLlepTB1}
N_{B-L}^{\rm lep}(T_{B1})=N_{\Delta_1}^{\rm lep}(T_{B1})+N_{\Delta_{\tilde{1}}}^{\rm lep}(T_{B1}) \, ,
\ee
where
\bea
N_{\Delta_1}^{\rm lep}(T_{B1}) & = &
p_{21}\,p_{32}\,\ve_3\,\kappa(K_3)\,e^{-{3\pi\over 8}\,(K_1+K_2)} \\ \nonumber
& & + p_{21}\,\ve_2\,\k(K_2)\,e^{-{3\pi\over 8}\,K_1} \\ \nonumber
& & + p_{\tilde{2}_3 1}\,(1-p_{32})\,\ve_3\,\k(K_3)\,e^{-{3\pi\over 8}\,K_1} \\ \nonumber
& & + \ve_1\,\k(K_1)
\eea
and
\bea
N_{\Delta_{\tilde{1}}}^{\rm lep}(T_{B1}) & = &
(1-p_{21})\,[p_{32}\,\ve_3\,\kappa(K_3)\,e^{-{3\pi\over 8}\,K_2}+\ve_2\,\k(K_2)] \\ \nonumber
& & + (1-p_{{\tilde{2}}_3 1})\,(1-p_{32})\,\ve_3\,\k(K_3) \, .
\eea
This calculation can be easily extended to  the asymmetry produced from RH neutrino decays
in the other mass patterns.

%%%%%%%%%%%%%%%%%%%%%%%%%%%%%%%%%%%%%%%%%%%%%%%%%%%%%%%%%%%%%%%%%%%%%%%%%%%%%%%%%%%%%%%%

\end{document}